\documentclass[aps,prx,twocolumn,showpacs,amsmath,amssymb,floatfix,notitlepage,superscriptaddress,longbibliography]{revtex4-2}

\usepackage{graphicx}
    \graphicspath{{./Images/}}
\usepackage{xcolor}
\usepackage[colorlinks]{hyperref}
\usepackage{enumitem}
    \setlist[itemize]{noitemsep, topsep=0pt}
    \setlist[enumerate]{noitemsep, topsep=0pt}
\usepackage{verbatim}
\usepackage{mathtools,amssymb,amsmath}
\usepackage{bm}
\usepackage{yhmath}
\usepackage[inline]{asymptote}
\usepackage{cancel}
\usepackage{relsize}
\usepackage{array}
\usepackage{bm}
\usepackage{physics}
\usepackage{booktabs}
\usepackage{times}
\usepackage{multirow}

\usepackage[all]{hypcap}

\newcommand{\be}{\begin{equation}}
\newcommand{\ee}{\end{equation}}
\newcommand{\<}{\langle}
\renewcommand{\>}{\rangle}

\usepackage{color}

\renewcommand{\vec}[1]{{\bf #1}}
\usepackage[]{hyperref}
\hypersetup{
  colorlinks = true,
  urlcolor = blue,
  linkcolor = blue,
  citecolor = blue
}

\usepackage[normalem]{ulem}

\begin{document}

\title{Complexity of Fermionic
Dissipative Interactions and Applications to Quantum Computing}

\author{Oles Shtanko}
\affiliation{Joint Center for Quantum Information and Computer Science,
NIST/University of Maryland, College Park, Maryland 20742, USA}
\affiliation{Joint Quantum Institute, NIST/University of Maryland, College Park, Maryland 20742, USA}

\author{Abhinav Deshpande}
\affiliation{Joint Center for Quantum Information and Computer Science,
NIST/University of Maryland, College Park, Maryland 20742, USA}
\affiliation{Joint Quantum Institute, NIST/University of Maryland, College Park, Maryland 20742, USA}
\author{Paul S. Julienne}
\affiliation{Joint Quantum Institute, NIST/University of Maryland, College Park, Maryland 20742, USA}

\author{Alexey V. Gorshkov}
\affiliation{Joint Center for Quantum Information and Computer Science,
NIST/University of Maryland, College Park, Maryland 20742, USA}
\affiliation{Joint Quantum Institute, NIST/University of Maryland, College Park, Maryland 20742, USA}

\begin{abstract}
Interactions between particles are usually a resource for quantum computing, making quantum many-body systems intractable by any known classical algorithm. In contrast, noise is typically considered as being inimical to quantum many-body correlations, ultimately leading the system to a classically tractable state. This work shows that noise represented by two-body processes, such as pair loss, plays the same role as many-body interactions and makes otherwise classically simulable systems universal for quantum computing. We analyze such processes in detail and establish a complexity transition between simulable and nonsimulable systems as a function of a tuning parameter. We determine important classes of simulable and nonsimulable two-body dissipation. Finally, we show how using resonant dissipation in cold atoms can enhance the performance of two-qubit gates.
\end{abstract}
\maketitle

Understanding whether a particular quantum system is easy or hard to simulate from the perspective of classical computation is a crucial task serving several goals. The first goal, as a primary step of many numerical studies, is to find efficient classical algorithms describing the desired quantum phenomena. Another goal arises in quantum computing, where finding many-body systems lacking an effective classical description may be worthwhile for constructing quantum computation \cite{nielsen_chuang_2010} and simulation \cite{feynman1982simulating,lloyd1996universal} devices. The versatility of the classical simulability problem can be illustrated by considering the sampling problem for noninteracting and interacting fermions \cite{knill2001fermionic, valiant2002quantum, terhal2002classical, aaronson2013bosonsampling}. There are efficient classical algorithms to simulate fermions described by a quadratic Hamiltonian: the amplitudes of time-evolved many-body configurations are expressed by an efficiently computable analytical formula \cite{terhal2002classical,wimmer2012algorithm}. The existence of an efficient algorithm makes the free-fermion approximation a numerically accessible and valuable method with applications to condensed-matter systems. At the same time, simulating interacting fermions is believed to be classically intractable. Indeed, simulating general interacting fermions is as hard as simulating the output of a universal quantum computer \cite{Schuch2009computational}. A similar practical differentiation between easy and hard problems can be applied to other systems \cite{jozsa2013classical,fefferman2015power, bremner2016average,PhysRevLett.121.030501, Muraleedharan2018}.

\begin{figure}[t]
    \centering
    \includegraphics[width=0.45\textwidth]{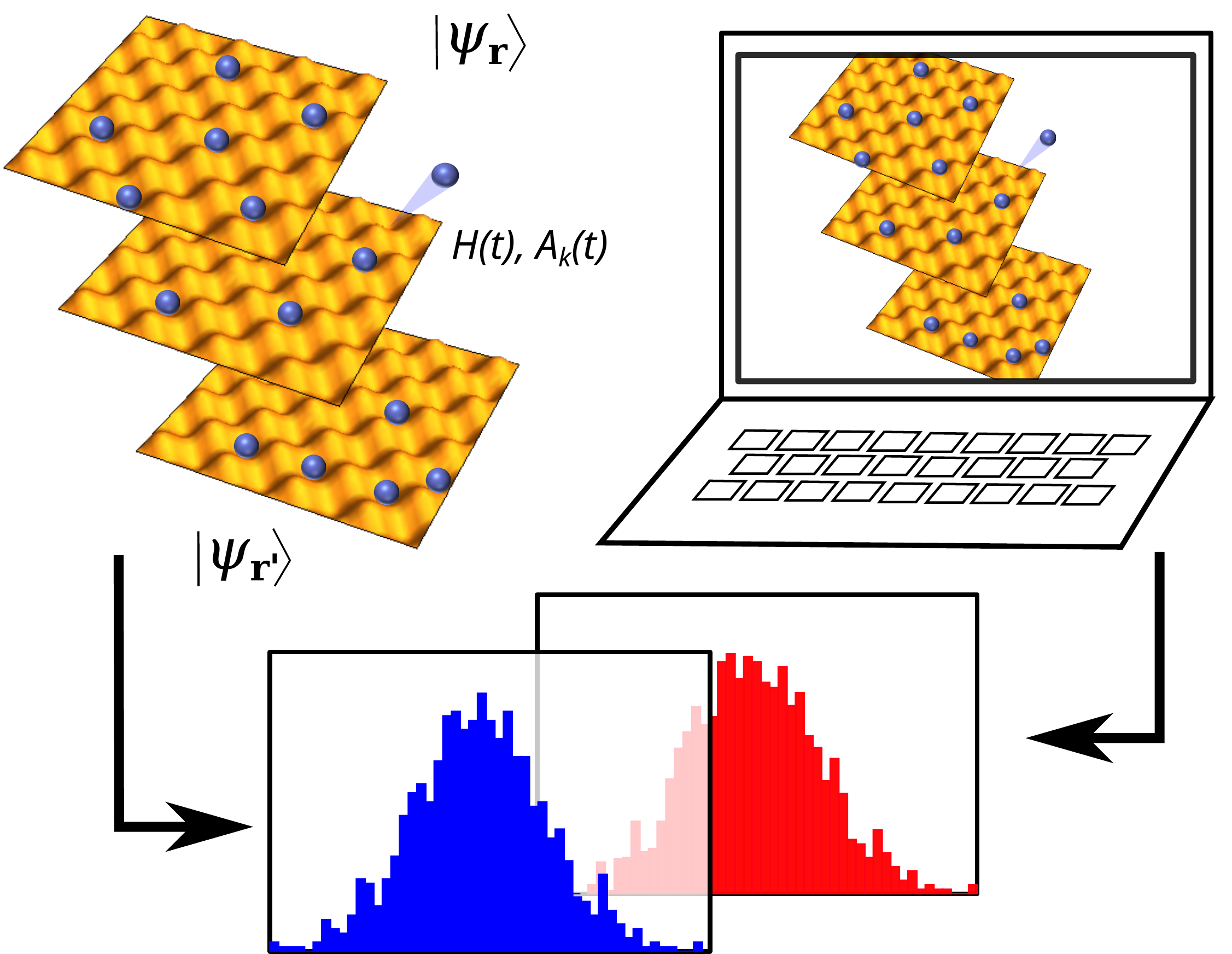}
    \caption{\textbf{Classical simulability}. We look for the existence of an efficient algorithm running on a classical computer and producing (sampling) the many-body configurations with the probability distribution close to the physical system after measurement in some basis.  We show that, for fermionic systems with Hamiltonian $H(t)$ and with dissipation described by quadratic Lindblad jump operators $A_k(t)$, such an algorithm exists for at least a restricted number of problems, while the worst-case scenario requires a quantum computer in order to be solved efficiently. The three optical lattices illustrate the state of the system at initial, intermediate, and final times.}
    \label{fig1}
\end{figure}

In this work, we study the fate of classical simulability of fermionic systems in the presence of dissipation, both for computing local observables and for sampling from the many-body output distribution (to be defined shortly). To obtain a classification of the complexity of simulating free fermions with dissipation, we consider a general class of Markovian processes, i.e.\ dynamics that depends only on the instantaneous system state and is independent of the preceding evolution \cite{breuer2002theory}. In previous studies, it was shown that Markovian single-fermion loss or gain terms keep the noninteracting system classically tractable \cite{prosen2010spectral, bravyi2011classical}.  As a step forward, we consider a much wider class of \textit{quadratic-linear} Lindblad jump operators.  Using the method of stochastic trajectories \cite{gardiner2004quantum,chenu2017quantum}, we establish a wide subclass of problems that can be simulated classically. At the same time, we demonstrate that, in general, quadratic Lindblad jump operators are at least as hard to simulate as most unitary interacting systems. More precisely, we establish a connection between dissipative interactions and fault-tolerant universal quantum computation exploiting the quantum Zeno effect \cite{misra1977zeno, degasperis1974does, franson2004quantum, sun2013photonic, daley2008quantum}. Therefore, evolution under quadratic Lindblad jump operators is equivalent in power to quantum computation. This effect can be compared with parity measurements, which can also make free-fermion dynamics universal \cite{beenakker2004charge}. The tractability and intractability results together show that simulation of { quartic dissipative Liouvillian operators} is a problem whose complexity can be changed from hard to easy by varying one or more parameters in the system \cite{PhysRevLett.121.030501}.

One motivation behind this work is the existence of a variety of accessible fermionic physical systems involving inelastic processes. Examples of dissipative processes described by quadratic Lindblad jump operators include two-body loss in trapped alkali atoms \cite{anderlini2007controlled,lee2007sublattice,zhang2006manipulation}, alkaline-earth atoms \cite{ludlow2011cold, zhang2015orbital,riegger2018localized, Scazza2014observation,targat2006accurate,cappellini2019coherent}, and cold molecules \cite{syassen2008strong, yan2013observation}. As we show, Feshbach resonances \cite{chin2010feshbach,Giorgini2008} can be used to significantly suppress coherent interactions between cold atoms, simultaneously increasing the rate of atom-pair trap losses. More general types of dissipation can be created by adding a source of atoms \cite{barontini2013controlling, labouvie2015negative, labouvie2016bistability} or inelastic photon scattering \cite{luschen2017signatures, patil2015measurement, li2019obesrvation}. In solid-state physics, examples of processes described by quadratic jump operators in Markovian approximation are Cooper-pair loss \cite{bonifacio1984superradiance,cassidy2017demonstration} and phonon-induced dephasing \cite{dolgirev2020nongaussian}.
Recent progress in the control of dissipative electronic systems has brought them into focus in condensed-matter physics. Some of the novel effects in noninteracting and mean-field fermionic systems include dissipation-induced magnetism \cite{lee2014heralded, lee2013unconventional, joshi2013quantum}, dissipative superfluids and superconductors \cite{diehl2010dissipation, yamamoto2019theory}, dissipative Kondo effect \cite{lourenco2018kondo, nakagawa2018nonhermitian}, non-Hermitian topological phases \cite{rudner2009topological, lee2016anomalous, leykam2017edge, yao2018edge, gong2018topological, shen2018topological, kawabata2019symmetry,Yoshida2019nonhermitian}, and non-Hermitian localization \cite{efetov1997directed, goldsheid1998distribution, longhi2015nonhermitian}.

We provide a classification of dissipative fermionic processes into \emph{easy} (efficiently simulable) and \emph{hard} (not efficiently simulable) classes according to their worst-case computational complexity. The classical simulability problem may be phrased in two ways, either in terms of evaluation of few-body observables or sampling from the full probability distribution on many-body outcomes. In the first task (\emph{few-body observables}), a classical computer is required to output the expectation value of an observable supported on $k$ sites, where $k$ does not grow with the system size. In the second task (\emph{sampling}), a classical computer is tasked with producing samples from the same distribution as the one obtained by measuring the time-evolved state in some canonical basis (see Fig.\,\ref{fig1}). Both tasks allow for the computer to make a small error $\epsilon$, measured appropriately in each case \footnote{To be precise, for the first task (estimating few-body observables), the error is measured by the maximum difference in the estimated and ideal expectation values of a unit-norm observable $O$. For the second task (sampling), the error is measured through the maximum variation distance between the ideal and the sampled probability distributions.}. The task of sampling is computationally harder; an algorithm producing samples in some product-state basis can also be used to obtain expectation values of few-body observables in the same basis. Therefore, in this work, we focus mainly on the easiness of sampling in arbitrary product-state bases as a criterion for overall easiness and on the hardness of computing few-body observables as a criterion for overall hardness. This choice gives the stronger of the two results for both easiness and hardness. 

We note that a limited version of classical simulability for some models below was also studied in previous works \cite{horstmann2013noise,vznidarivc2010exact,eisler2011crossover}. In particular, it was shown that two-point correlation functions in such models can be evaluated via solving a closed set of equations. This result establishes classical simulability of local observables and can be used in various problems such as dissipative transport or optical response. However, this result alone is not sufficient to establish the simulability of sampling. In contrast to local observables, simulated sampling requires the full knowledge of the many-body output probabilities, therefore the sampling complexity of systems with simulable low-order correlations remains unclear. As we revisit below, Gaussian systems are the exception that allow reducing these output probabilities to two-point correlation functions via Wick’s theorem; other systems we study below do not have such simple reduction (see Appendix \ref{app:non-gauss}). To overcome this problem, we develop the easiness proof that does not require applicability of Wick’s theorem. In conclusion, sampling is a stronger notion of simulation compared to two-point correlators in a sense that any local observables can be efficiently obtained using an oracle producing sampling outcomes

Let us emphasize the importance of the provided complexity analysis. While established easy dissipation classes are limited to certain fine-tuned processes, such limited simulable models have an important application for quantum computing. For example, classical models can be used in calibration of quantum computers, simulation of the impact of noise on sampling, and analysis of fermionic quantum error-correcting codes \cite{bravyi2010majorana}. More fundamentally, identifying easy classes is an important first step to analyze easy-hard transitions in open fermionic systems, as we analyze below. 
At the same time, the hardness result we obtain in this work establishes utilizing dissipative interactions as an alternative path toward building a universal quantum computer. This conclusion is surprising, since dissipative interactions generally produce mixed states. However, dissipative interactions can be used only in a manner utilizing a blockade mechanism induced by the quantum Zeno effect, as we show below. In cold atomic systems, controlling dissipative interactions differs from photonic systems studied before \cite{franson2004quantum, sun2013photonic} and can be achieved using an atomic Feshbach resonance. In this paper, we analyze in detail a scheme for universal quantum computation with $^{40}$K atoms and illustrate that, with realistic experimental parameters, an entangling gate with low error rate of roughly $10^{-4}$ is possible. Existence of both easy and hard classes for two-body dissipation establishes it as a valuable model for physical analysis of noisy intermediate-scale quantum devices.

\textit{Model.---} We consider dynamics generated by the Lindblad master equation \cite{Lindblad1976,breuer2002theory}
\be\label{eq:Lindblad_equation}
\begin{split}
\frac {d\rho}{dt} 
&= -i[H(t), \rho]+\sum_{k=1}^{k_{A}}A_k(t)  \rho  A^\dag_k(t)-\frac 12\{A^\dag_k(t) A_k(t), \rho \},
\end{split}
\ee
where $\{X,Y\} \equiv XY+YX$ is the anticommutator, $\rho(t)$ is the density matrix  of the system, $H(t)$ is a noninteracting Hamiltonian, and $A_k(t)\in\mathcal A(t)$ form a set of $k_A$ Lindblad jump operators. We set $\hbar = 1$ throughout the paper unless specified otherwise. Both the Hamiltonian and the Lindblad jump operators may depend explicitly on the time but not on the state itself. The corresponding map $\rho(t_2)  = \mathcal V(t_2,t_1)\rho(t_1)$ between arbitrary times $t_1$ and $t_2\geq t_1$ satisfies $\mathcal V(t_2,t_1) = \mathcal V(t_2,\tau) \mathcal V(\tau,t_1)$ for any $t_2\geq\tau\geq t_1$. This divisibility condition is commonly referred to as the most general definition of Markovian dynamics \cite{PhysRevA.81.062115}.  The master equation in Eq.~\eqref{eq:Lindblad_equation} is invariant under certain transformations of the set of Lindblad jump operators $\mathcal A(t)$, such as operator permutations, multiplying any Lindblad operator by a phase factor, or splitting and merging of identical operators as $A_k \rightleftarrows \{\sqrt{p}A_k, \sqrt{1-p}A_k\}$, $0\leq p\leq 1$.

As a physical system of interest, we consider a fermionic many-body problem where $N\leq L$ spinless fermions initially occupy $L$ available levels. Such systems are commonly described by the second quantization method, which expresses any operator, including the Hamiltonian and Lindblad jump operators, in terms of fermionic Fock operators $c^\dag_n$ and $c_n$, $n \in \{0,1,\ldots L-1\}$. Fock operators create and annihilate a single fermion in a particular mode and satisfy the canonical commutation relations $\{c_n,c_m\}=0$, $\{c^\dag_n,c_m\}=\delta_{nm}$. Though the conventional fermion operators are suitable in most physical problems, in the absence of fermion number conservation it is convenient to use the $2L$ Hermitian Majorana fermion operators $\gamma_{2n} = c_n+c^\dag_n$ and  $\gamma_{2n+1} = -i(c_n-c^\dag_n)$, due to their simple anticommutation relations $\{\gamma_i,\gamma_j\} = 2\delta_{ij}$, $i,j\in \{0,1,\dots,2L-1\}$. We consider the most general form of a noninteracting Hamiltonian \cite{colpa1979diagonalisation}
\be\label{eq: hamiltonian}
H(t) = \frac i2 \sum_{i,j=0}^{2L-1} \alpha_{ij}(t)\gamma_{i}\gamma_j+\sum_{i=0}^{2L-1} \beta_i(t)\gamma_i,
\ee
where $a_k(t)$ are antisymmetric $2L\times 2L$ matrices, $b_k(t)$ are $2L$ vectors, and $d_k(t)$ are numbers; all the parameters are complex-valued in general.We assume that the magnitude of all entries of $\alpha(t)$ and $\beta(t)$ and their time derivatives scale at most polynomially with system size.

\begin{table}[]
\begin{tabular}{@{}llr@{}} \toprule
Type & Examples of $A_k$  & Complexity \\ \midrule
Dephasing & $c^\dag_1c_1$ & \multirow{4}{*}[-2pt]{Easy (EC1)} \\
Particle shuffle &$c^\dag_1c^{}_2$ \& $c^\dag_2c^{}_1$  & \\
Classical fluctuations & $c^\dag_1$ \& $c^{}_1$  &  \\
Classical pair fluctuations & $c^\dag_1c^\dag_2$ \& $c^{}_1c^{}_2$  &  \\ \midrule
Mixing unitaries & $2c^\dag_1c^{}_1-1 +i(c^\dag_2+c^{}_2)$&  Easy (EC2)\\ \midrule
Single-particle loss/gain & $c_1$  {\rm OR } $c_1^\dag$ & Easy (EC3) \\ \midrule
Incoherent hopping & $c^\dag_1c_2$& \multirow{2}{*}[-2pt]{Hard} \\
Pair loss/gain & $c_1c_1$ {\rm OR } $c^\dag_1c^\dag_1$ & \\ \bottomrule
\end{tabular}
\caption{Comparison between different types of noninteracting fermion dynamics with additional dissipation. For simplicity, we provide examples for two modes out of $L$, denoted by numbers $1$ and $2$. The symbol \& means that both operators are present in the set $\mathcal A(t)$ with factors equal in absolute value. Abbreviations EC1, EC2, EC3 stand for easy class 1, 2, and 3 described in the text. \textit{Note:} Different classes can be combined by summing the right-hand sides of the corresponding master equations. However, the jump operators cannot be combined without precaution: the sum of two easy jump operators from different classes does not necessarily produce a simulable jump operator. }
\label{tab:summary}
\end{table}

In this work, we focus on the classical resources needed to approximately sample from the fermion distribution at time $t$,
\be\label{eq:sampling_prob}
P(\vec r|\vec r') = \<\psi_{\vec r}|\rho(t)|\psi_{\vec r}\>, \quad \rho(0) = |\psi_{\vec r'}\>\<\psi_{\vec r'}|,
\ee
where the vectors $\vec r'$ and $\vec r$ denote the positions of occupied modes in the initial and final (measured) product-state configurations, respectively, and $|\psi_{\vec r}\>$ is a product state defined as
$|\psi_{\vec r}\> = c^\dag_{r_1}\dots c^\dag_{r_N}|0\> = \gamma_{2r_1}\dots\gamma_{2r_{N}}|0\>$. Here $|0\>$ is the vacuum state defined as the state satisfying $c_n|0\> = 0$ for all $n$.
Importantly, because the dynamics may not conserve the total fermion number, the final number of fermions $\tilde N$ can, in general, be different from the initial number: $N \neq \tilde N$.

We establish the sufficiency of polynomial resources for classically simulating dynamics due to arbitrary noninteracting Hamiltonians in Eq.~\eqref{eq: hamiltonian} and a limited set of Lindblad jump operators $A_k(t)\in \mathcal A(t)$ in the worst case. In order to prove polynomial-time simulability (also called easiness) for limited classes of dissipative dynamics, we reduce the problem to that of simulating unitary noninteracting fermionic evolution, an easy problem for a classical computer. In order to prove hardness for more general Lindblad jump operators, we  exploit the ability of dissipative dynamics to perform arbitrary quantum computation (i.e.\ we prove that simulating universal quantum computation reduces to simulating Lindbladian dynamics). 
 
The results of this work are briefly illustrated in Table \ref{tab:summary}. First of all, we define three classically tractable classes of Lindblad jump operators (defined as easy classes 1, 2, and 3). All of these cases allow for polynomial-time sampling of any Hamiltonian and Lindblad jump operators from the given class on a classical computer, with error scaling inverse-polynomially with $L$. Easy Class 1 (EC1) allows for simulation of self-adjoint sets of quadratic Lindblad jump operators: all Lindblad jump operators in the set $\mathcal A(t)$ come with their Hermitian conjugate. This class includes such widely used examples as dephasing, incoherent particle shuffle, and classical fluctuations of the number of fermions and of the number of fermion pairs. Easy class 2 (EC2) works with unitary quadratic Lindblad jump operators.  Finally, easy class 3  (EC3) describes the loss or gain of a single particle in the system and can be used in combination with EC1 and/or EC2. At the same time, there exists a class of Lindblad jump operators with a nonzero measure that is hard to classically simulate. Examples from this class include pair loss or gain and incoherent fermion hopping. Below we explore each class separately.

We focus on quadratic-linear Lindblad jump operators of the form
\be\label{eq:single_body_A}
A_k(t) = \frac i2 \sum_{i,j=0}^{2L-1} a_k^{ij}(t)\gamma_{i}\gamma_j + \sum_{i=0}^{2L-1} b_k^i(t)\gamma_i+d_k(t),
\ee
where $a_k(t)$ and $b_k(t)$ are arbitrary complex-valued $2L \times 2L$ matrices and $2L$-vectors respectively, and $d_k(t)$ is a number. Notably, the same master equation allows representation with more than one set of jump operators $\mathcal A = \{A_k\}$. A set can be reduced to a smaller one if its jump operators are linearly dependent \cite{breuer2002theory}.  Therefore, the number $k_A$ of Lindblad jump operators in the smallest set does not exceed the number of linearly independent quadratic operators, i.e. $L(L+1)$. 
Also, as with the Hamiltonian, we assume that the magnitude of the entries of $a_k(t),b_k(t)$, and $d_k(t)$ and their time derivatives grow at most polynomially with the system size. 

This work is organized as follows. In section \ref{sec:free}, we provide a brief introduction to free-fermion sampling, recalling established results in the literature and connecting them to the most general form of quadratic-linear Hamiltonians. In Section \ref{sec:easy}, we derive three new algorithms allowing us to solve distinct classes of fermionic problems involving quadratic Lindblad jump operators and prove that these algorithms run in time that is polynomial in both $L$ and the inverse of the distance from the exact distribution.
In Section \ref{sec:hard}, we establish generic class of systems that belong to the hard class and show their robustness to the presence of minor imperfections. 
In Section \ref{sec:atoms}, we show how our complexity result applies to realistic experimental settings. We demonstrate that natural pair loss in cold atomic systems can be controlled and utilized to implement universal quantum computing. The dissipation-assisted gates provide an alternative to unitary gates with a potential advantage in the speed of two-qubit operations.

\section{Free-fermion sampling \label{sec:free}}

In this Section, we discuss the noninteracting fermion problem in the absence of dissipation. We recap the work of Terhal and DiVincenzo \cite{terhal2002classical}, which shows that all output probabilities $P(\vec r|\vec r')$ in Eq.~\eqref{eq:sampling_prob} and the marginal probabilities can be obtained using a classically tractable analytical formula. Before referring to this result, we need to incorporate the linear terms present in Eq.~\eqref{eq: hamiltonian} into effective quadratic dynamics. In order to do so, we consider a slightly larger system containing an extra ancilla $(L+1)$th mode \cite{colpa1979diagonalisation}, labeled as $n=L$. Next, we choose new effective dynamics such that the ancilla mode remains in the state $|+\> \equiv (|0\>+|1\>)/\sqrt{2}$ during the entire evolution, including the initial and final times, i.e.\
\be
|\psi_{\vec r'}\>\to|\psi_{\vec r'}\>\otimes|+\>, \qquad |\psi_\vec r\>\to|\psi_\vec r\>\otimes|+\>.
\ee
To construct such dynamics, we consider a new Hamiltonian by replacing $\gamma_i \to i\gamma_i\gamma_{2L}$, where $\gamma_{2L}$ and $\gamma_{2L+1}$ are Majorana operators acting on the ancilla mode. It is straightforward to check that such a transformation results in a new purely quadratic Hamiltonian (without any linear terms) that keeps the state of the ancilla stationary and does not modify the dynamics of the original Hamiltonian. The new coefficients in Eq.~\eqref{eq: hamiltonian} are 
\be
\alpha_{ij} \to \tilde\alpha_{ij} = \alpha_{ij}+\delta_{i2L}\beta_{j}-\delta_{j2L}\beta_{i},
\ee
where we use by default $\beta_{2L}=\beta_{2L+1} = 0$.
Given that the modified initial and final conditions for the system and the ancilla are $\{\vec r'\}\to \{\vec r',s'\}$, $\{\vec r\}\to \{\vec r,s\}$, $s,s' \in \{0,1\}$, the probability $P(\vec r|\vec r')$ of obtaining  outcome $\vec{r}$ for the original system can be computed from the probability $P(\{\vec r,s\}|\{\vec r',s'\})$ for the system with the ancilla as follows:
\be
P(\vec r|\vec r') =  \frac 12\sum_{s,s'\in\{0,1\}}P(\{\vec r,s\}|\{\vec r',s'\}).
\ee
Summarizing, this method ensures that the dynamics of a linear-quadratic Hamiltonian can always be reduced to the dynamics of a quadratic one by expanding the system size by one mode. Therefore, we henceforth consider only quadratic Hamiltonians.

Let us derive the formula for the sampling probability. We start from a (backwards) time-evolved Majorana fermion operator $\gamma_i(t) = U_t \gamma_iU_t^\dag$, where $ U_t= \mathcal T\exp\left(-i\int_0^t H(t')dt'\right)$. Here $\mathcal T\exp$ is the standard time-ordered exponential.
Given the quadratic structure of the Hamiltonian, this evolution is a linear transformation $
\gamma_i(t)= \sum_i R_{ij}(t)\gamma_j$, 
where $R = \mathcal T\exp\left(- 2\int_0^t\alpha(t')dt'\right)$ is a unitary $2L\times2L$ matrix. One can use this expression to derive the time evolution of a fermion operator as
\be\label{eq:c_transform}
U_t c_nU_t^\dag = \frac 12 U_t(\gamma_{2n}+i\gamma_{2n+1})U_t^\dag = \sum_j T_{nj}\gamma_j,
\ee
where $T_{nj} \equiv R_{2n,j}+iR_{2n+1,j}$ are elements of a $L\times 2L$ transformation matrix $T$. Labeling the initially empty sites as $l'_i$ and recalling that the initial fermion positions are $r'_i$ and that the final positions are $r_i$, the linearity allows to write the output probability in Eq.~\eqref{eq:sampling_prob} at any time as
\be\label{eq:sampling_T}
\begin{split}
&P(\vec r|\vec r')= \bra{\psi_{\vec r}} U_t \ketbra{\psi_{\vec r'}} U_t^\dag \ket{\psi_{\vec r}}\\
&= \<\psi_\vec r|U_t  c^\dag_{r'_1} c_{r'_1}\dots  c^{}_{l'_{L-N}}c^\dag_{l'_{L-N}}U_t^\dag |\psi_\vec r\>  \\
&=\sum_{n_1,\ldots n_L;m_1,\ldots m_{L-N}} T^*_{r_1'n_1} T_{r_1'm_1}\dots T_{l'_{L-N}m_L} T^*_{l'_{L-N}n_L}\times\\
&\qquad\<0|\gamma^{}_{2r_{N}}\dots \gamma^{}_{2r_1}\gamma_{n_1}\gamma_{m_1}\dots\gamma_{m_L}\gamma_{n_L} \gamma^{}_{2r_1}\dots\gamma^{}_{2r_{N}}|0\>.
\end{split}
\ee

This expression can be computed efficiently using Wick's theorem and written in a compact form.  Let $\mathcal I$ be a subset of indices with increasing order and $A[\mathcal I]$ be the matrix whose elements satisfy  $A[\mathcal I]_{ij}\equiv A_{\mathcal I_i,\mathcal I_j}$. Consider the set $\mathcal I = \{r'_i,L+ l'_j,2L+2 r_k\}$, where $i\in \{1,2,\ldots N\}$, $j\in \{1,2,\ldots L-N\}$, and $k \in \{1,2,\ldots \tilde N\}$ take all possible values. Then the result can be written as \cite{terhal2002classical}
\be\label{eqm:unitary_sampling_probability}
P(\vec r|\vec r') = {\rm Pf}\,M[\mathcal I],
\ee
where $\rm Pf$ is the Pfaffian, and $M$ is a $4L\times4L$ matrix
\be
M =
\left(\begin{matrix}
T\Lambda T^T  & T\Lambda T^\dag & T\Lambda \\
T^*\Lambda T^T & T^*\Lambda T^\dag & T^*\Lambda\\ 
\Lambda T^T & \Lambda T^\dag & I\\ 
\end{matrix}\right),
\ee
where, in turn, the $2L\times 2L$ matrix $\Lambda$ is
\be
\Lambda = 
I_{L\times L}\otimes\left(\begin{matrix}
1 &i\\
-i&1
\end{matrix}\right).
\ee
%Because matrix $M$ linearly scales with the system size, 
The expression in Eq.~\eqref{eqm:unitary_sampling_probability} can be efficiently evaluated on a classical computer using existing polynomial-time algorithms for computing Pfaffians \cite{wimmer2012algorithm}. Similarly, marginal probabilities can be efficiently computed conditioning on the output of a given fraction of sites, as in Ref.\,\cite{terhal2002classical}, which is enough to efficiently sample from the output probability distribution.

\section{Easy Classes \label{sec:easy}}

Here we present three algorithms that allow simulating specific fermionic problems involving quadratic Hamiltonians and quadratic-linear Lindblad jump operators.
All methods are based on stochastic unraveling, i.e., replacing dissipative dynamics by a stochastic free-fermion Hamiltonian without changing the outcome distribution { (see also Ref.~\cite{chenu2017quantum})}. 
Since each stochastic realization can be simulated efficiently by a classical computer, as established in the previous section, a classical computer can serve as a black-box sampler that reproduces measured outcomes. 
In this section, we demonstrate that the classes of problems belonging to the aforementioned easy classes 1--3 are efficiently simulable.
In particular, we show that these problems require computation resources $C$ (number of operations) bounded as $C \leq {\rm poly}\left(L,t^2/\epsilon\right)$ to sample from a distribution that is $\epsilon$-close to the target distribution. 
Therefore, we establish efficient classical algorithms for approximate dissipative fermion sampling in the presence of certain classes of quadratic-linear Lindblad jump operators.

\subsection{Efficient classical algorithms}

Let us define \textit{Easy Class 1} (EC1) as problems that involve quadratic-linear self-adjoint Lindblad sets $\mathcal A(t)$ defined as follows. We assume that at any time one can divide the set as a union of two equal-size subsets, $\mathcal A = \mathcal{A}_1 \cup \mathcal{A}_2$, where the Hermitian conjugate of every Lindblad operator in $\mathcal{A}_1$ returns an operator in $\mathcal{A}_2$ (and vice versa). Under this division, any Hermitian Lindblad operator must be included in both subsets $\mathcal{A}_1$ and $\mathcal{A}_2$ with normalization factor $1/\sqrt{2}$. The latter splitting can be seen as a transformation that keeps the Lindblad equation invariant, as defined earlier below Eq.~\eqref{eq:Lindblad_equation}. Examples from EC1 include several important physical models such as dephasing and classical fluctuations (see examples of sets in lines 1--4 in Table \ref{tab:summary}).

In previous works, it was shown that such systems have two-point correlation functions that are classically simulable by solving a closed set of linear equations \cite{horstmann2013noise,vznidarivc2010exact,eisler2011crossover}. This is indeed a strong indication that the system can be simulable in the broader context of sampling complexity. However, as we noted previously, Wick's theorem is not applicable to non-Gaussian states (see Appendix \ref{app:non-gauss}). This means that the scheme we utilized to obtain Eq.~\eqref{eqm:unitary_sampling_probability} does not work any more. We now show an alternative scheme using stochastic unraveling that leads us to the easiness result.

To efficiently simulate dynamics from EC1, we consider a stochastic Hamiltonian
\be\label{eq:stoch_ham1}
H'(t) = H(t)+\sum_{A_k \in \mathcal{A}_1} \theta_{k}(t)A_k(t)+\theta^\dag_{k}(t)A^\dag_k(t),
\ee
where $\theta_{k}(t)$ is a complex random variable taking constant values $\theta_{k}(t) = \xi_{nk}/\sqrt{\Delta\tau}$ during short time intervals $t \in [n\Delta \tau, (n+1)\Delta \tau]$. Later we also consider $\theta_k(t)$ as operators.
The discrete complex Gaussian variables $\xi_{nk}$ satisfy $\mathbb E \xi_{nk}=0$, $\mathbb E\xi^*_{nk}\xi_{n'k'}=\delta_{nn'}\delta_{kk'}\delta_{ab}$, where $\mathbb E$ denotes the expectation value taken over the random variables. 
Then, given a stochastic Hamiltonian of the form in Eq.\,\eqref{eq:stoch_ham1}, the original dynamical map $\mathcal V(t_2,t_1)$ generated by the Lindblad equation can be 
approximated as
\be\label{eqm:Liouville_conv}
\mathcal V(t_2,t_1) = \mathbb E \mathcal V_{\rm st}(t_2,t_1) + \delta\mathcal V (t_2,t_1)\Delta\tau+O(\Delta\tau^2),
\ee
where $\delta\mathcal V (t_2,t_1)\Delta\tau$ is the lowest-order correction (to be explicitly derived below) and $\mathcal V_{\rm st}$ is a stochastic unitary map
\be\label{eq:stoch_unitary_map}
\mathcal V_{\rm st}(t_2,t_1)\rho = U(t_2,t_1)\rho U^\dag(t_2,t_1).
\ee
In the above, $U(t_2,t_1) = \mathcal T\exp(-i \int_{t_1}^{t_2} dt'H'(t'))$ encodes the time evolution due to $H'(t)$ in Eq.~\eqref{eq:stoch_ham1}. The average $\mathbb E$ in Eq.~\eqref{eqm:Liouville_conv} is taken  over the stochastic fields $\theta_k(t)$.
The resulting output probabilities satisfy
\be\label{eq:avg_prob}
P(\vec r|\vec r') =  \mathbb E \,P_{\rm st}(\vec r|\vec r')+O(\Delta\tau),
\ee
where $P_{\rm st}(\vec r|\vec r')$ is the output probability for the unitary dynamics in Eq.~\eqref{eq:stoch_unitary_map} obtained via the formula in  Eq.~\eqref{eqm:unitary_sampling_probability}. Therefore, a computer programmed to sample from the distribution for a randomly chosen set of unitary trajectories will produce outcomes with the same probabilities as the physical process following Lindbladian evolution, up to $O(\Delta\tau)$ error. The cost of suppression of this error in terms of computational resources will be discussed later in this section. Here we just specify that the correction to the dynamical map, which we treat as an error, can be expressed as
\be\label{eq:V_sp_decomposition2}
\begin{split}
\delta\mathcal V(t_2,t_1) = \mathbb E\int_{t_1}^{t_2}dt'\mathcal V_{\rm st}(t_2,t') \mathcal D(t')\mathcal V_{\rm st}(t',t_1),
\end{split}
\ee
where $\mathcal D(t)$ is a time-local superoperator
\be\label{eq:correction_form}
\mathcal D(t)\rho = \sum_\alpha D^{(1)}_{\alpha}(t)\rho D^{(2)}_{\alpha}(t).
\ee
Here, the operators $ D^{(i)}_{\alpha}(t) = {\rm poly}_4(H(t),A_k(t))$ can be expressed as polynomials of degree four in terms of the Hamiltonian and Lindblad jump operators at time $t$. Therefore, $ D^{(i)}_{\alpha}(t)$ can always be presented as a sum of terms, each being a product of no more than eight Majorana operators. The specific form of these operators and the derivation of Eq.~\eqref{eqm:Liouville_conv} can be found in Appendix \ref{app:ec1}.

{ Although the proposed unraveling scheme represents dynamics in terms of stochastic trajectories for Gaussian pure states, the resulting averaged mixed state is non-Gaussian, in contrast to previously studied problems \cite{prosen2010spectral, bravyi2011classical}. Therefore, the overall dynamics of EC1 represents dissipative interactions of fermions, while the proposed method can be seen as a good choice of time-dependent density matrix decomposition.}

Let us consider another class of problems, \textit{easy class 2} (EC2), that include unitary quadratic Lindblad jump operators $A_k = \sqrt{\Gamma_k(t)}Y_k(t)$, where $\Gamma_k(t)\geq0$ are time-dependent rates and $Y_k(t) = \exp(-iG_k(t))$ are unitary operators generated by quadratic-linear Hamiltonians $G_k(t)$ of the form in Eq.~\eqref{eq: hamiltonian}. 
To classically simulate dynamics under EC2, we also consider discretized time evolution with sufficiently small timesteps $\Delta \tau$ and set the unitary transformation $U(t_1,t_2) = \prod_{n=n_1}^{n_2} U_n$, where the timestep unitaries $U_n$ are generated randomly according to the rule
\be\label{eq:stochastic_unitary}
U_n = U_n^0\times
\begin{cases}
Y_k(t_n),\quad & p_k =\Gamma_k(t_n) \Delta \tau, \\
I,\quad &p_0 = 1-\sum_k\Gamma_k(t_n) \Delta \tau.
\end{cases}
\ee
Here $p_k$ are probabilities that are used to generate the respective outcomes, $U_n^0 = \mathcal T \exp(-i \int_{n\Delta\tau}^{(n+1)\Delta\tau} H(t)dt)$, and $t_n\in [n\Delta\tau,(n+1)\Delta\tau]$ are random times generated from the uniform distribution. 

Notwithstanding the slightly different stochastic unraveling, the procedure for approximating EC2 is the same as for EC1. In particular, the system dynamics is described by the expression in Eq.~\eqref{eqm:Liouville_conv} leading to the distribution in Eq.\,\eqref{eq:avg_prob}, with the average taken over stochastic unitary realizations. The correction term has the form in Eq.~\eqref{eq:correction_form}, but the operators $D^{(i)}_{\alpha}(t)$ here are degree-two polynomials in the Hamiltonian and Lindblad jump operators. The detailed form of the operators along with the derivation can be found in Appendix \ref{app:ec2}.

Finally, let us consider \textit{easy class 3} described by generic linear Lindblad jump operators $A_k(t) = \sum_i b_k^i\gamma_i+d_k$, which can be obtained by setting $a_k=0$ in Eq.~\eqref{eq:single_body_A} without assuming any additional restrictions on the set $\mathcal A(t)$. Previous works had already shown that linear jump operators can be simulated classically \cite{prosen2010spectral, bravyi2011classical}. However, this proof applies only to Gaussian states and
cannot be extended to, for example, Lindblad equations that also contain easy quadratic jump operators. Below we propose a different way of simulating linear jump operators similar to the one we used for EC1. This technique would allow us to combine EC1, EC2, and EC3 into a single easy class of Lindblad equations, including both quadratic and linear processes.

%The simulation of this class uses the same method as EC1 but requires a presence of
Now let us show that simulation of linear jump operators is equivalent to simulating a unitary system extended by a number of ancilla modes. In particular, we require $L_a = t/\Delta\tau$ ancilla fermion modes equal to the number of time steps after discretization. Let us enumerate the ancilla modes described by Majorana fermion operators $\gamma_{2n}$ and $\gamma_{2n+1}$ using indices $n=L,\dots, L+L_a-1$. We also assume that the ancilla modes are initialized in the vacuum state and traced out after performing the evolution. 
The dynamics of both the system and the ancillas can be described as unitary evolution with the  Hamiltonian in Eq.~\eqref{eq:stoch_ham1}, with one important difference.
Now, the quantities $\theta_{k}(t)$ in the time interval $t \in [n\Delta\tau, (n+1)\Delta\tau]$ are operators instead of numbers, and are given by
\be\label{eq:ec3_stoch_field}
\begin{split}
\theta_{k}(t) = \xi_{nk}\Delta\tau^{-1/2}(\gamma_{2(L+n)}+i\gamma_{2(L+n)}).
\end{split}
\ee
The random variables $\xi_{nk}$ are the same as in EC1.
The idea is that we pair a loss (gain) term on the system with a gain (loss) term on the ancilla to make the overall Hamiltonian quadratic. 
After discarding the ancilla modes, the evolution becomes equivalent to the target dissipative dynamics, up to a discretization error that originates from the approximation in Eq.~\eqref{eqm:Liouville_conv} and Eq.~\eqref{eq:correction_form}, with $D^{(i)}_{\alpha}(t)$ expressed as degree-four polynomials in the Hamiltonian and Lindblad jump operators, as shown in Appendix \ref{app:ec3}.

We note that the stochastic simulation method takes advantage of the system state being a convex mixture of Gaussian \emph{density} operators \cite{oszmaniec2014classical,de2013power}. This is a particular case of the more general property that any state of a fermionic system with well-defined parity has a representation as a convex distribution over generic (not necessarily Hermitian) Gaussian operators \cite{corney2006gaussian}.

\subsection{Performance of the classical algorithms} 

Let us quantify the error of the method of quantum trajectories used for easy classes 1--3, and then show that the sampled distribution can be made arbitrarily close to the exact one with an appropriate choice of the timestep $\Delta \tau$. 
In order to characterize the approximation error $\epsilon$ associated with sampling from a distribution $\tilde P(\vec r|\vec r')$ different from the ideal distribution $P(\vec r|\vec r')$, we utilize the total variation distance
\be
\epsilon = \frac 12 \max_{\vec r'}\sum_{\vec r}\bigl|\tilde P(\vec r|\vec r')-P(\vec r|\vec r')\bigl|,
\ee
where the maximization is over all possible initial product-state configurations $\vec r'$.

Using convexity of the absolute value and the expression for the correction in Eqs.~\eqref{eq:V_sp_decomposition2}--\eqref{eq:correction_form}, the error can be bounded as (see Appendix \ref{app:error}),
\be\label{eq:erro_int_step1}
\epsilon \leq \frac {\Delta\tau}2 \max_{\vec r'}\sum_\alpha\int_0^t dt'C^\alpha_{\vec r'}(t,t')+O(\Delta\tau^2),
\ee
where 
\be\label{eq:C_term}
C^\alpha_{\vec r'}(t,t') =  \mathbb E\sum_{\vec r}\Bigl| \<\psi_{\vec r}| D^{(1)}_\alpha(t,t')\rho_{\vec r'}(t) D^{(2)}_{\alpha}(t,t')|\psi_{\vec r}\>\Bigl|.
\ee
Here $D_\alpha^{(i)}(t,t') = \mathcal V_{\rm st}(t,t') D_\alpha^{(i)}(t')$ and $\rho_{\vec r'}(t) = \mathcal V_{\rm st}(t,0)\rho_{\vec r'}$ are operators transformed according to unitary evolution for a single stochastic trajectory, and the average $\mathbb E$ is taken over all trajectories.
We now use the following lemma to further bound this expression.\\

\textbf{Lemma}. \textit{Consider two sparse operators $O_1$ and $O_2$ whose matrix elements satisfy
\be
\begin{split}
\<\psi_{\vec r}|O_\alpha|\psi_{\vec r'}\> = 0 \quad {\rm if}\quad d_H(\vec r,\vec r')\geq k_\alpha, \ \ \alpha\in\{1,2\},
\end{split}
\ee
where $d_H(\vec r,\vec r')$ is the Hamming distance between states with the positions of fermions $\vec r$ and $\vec r'$. Let $\rho$ be a normalized positive semidefinite operator, $\rho\geq0$, $\Tr \rho = 1$, then
\be\label{eq:lemma_statement}
\sum_{\rm \vec r} |\<\psi_{\vec r}|O_1\rho O_2|\psi_{\vec r}\>|\leq \frac {L^{k_1+k_2}}{k_1!k_2!}\|O_1\|_{\rm max}\|O_2\|_{\rm max},
\ee
where $\|O_\alpha\|_{\rm max}$ is the max-norm.}\\

The proof of the Lemma can be found in Appendix \ref{app:error}. The result of the Lemma allows us to simplify Eq.~\eqref{eq:C_term} as
\be\label{eq:c_as_porduct_of_Ds}
\begin{split}
C^\alpha_{\vec r'}(t,t')\leq \frac {L^{k_{1\alpha}+k_{2\alpha}}}{k_{1\alpha}!k_{2\alpha}!}\mathbb E \|D^{(1)}_\alpha(t,t')\|_{\rm max}\|D^{(2)}_\alpha(t,t')\|_{\rm max},
\end{split}
\ee
where $k_{i\alpha}$ is the locality of the operator $D^{(i)}_\alpha(t,t')$, i.e.\ the maximum number of Majorana operators in its decomposition. 
Because $\mathcal V_{\rm st}$ is a map describing free-fermion evolution, the locality $k_{i\alpha}$ of the operator $D^{(i)}_\alpha(t,t')$ is equal to the locality of $D^{(i)}_\alpha(t')$. 
At the same time, as analyzed in the previous section, the localities of operators $D^{(i)}_\alpha(t')$ satisfy $k_{\mu\alpha}\leq k_{m}$, where $k_{m} = 8$ for EC1/EC3, and $k_{m} = 4$ for EC2. 
We can also bound the max-norm by the (spectral) operator norm
\be\label{eq:bound_on_D}
\|D_\alpha^{(i)}(t,t')\|_{\rm max}\leq \|D_\alpha^{(i)}(t,t')\| = \|D_\alpha^{(i)}(t')\|.
\ee
As a result, the error bound is given by
\be\label{eq:error_bound_final}
\epsilon \leq \frac {\Delta\tau}2 \frac{L^{2k_{m}}}{(k_{m}!)^2}  \mathbb E \sum_\alpha\int_0^t dt'\|D_\alpha^{(1)}(t')\|\|D_\alpha^{(2)}(t')\| .
\ee
Since the matrices $D_\alpha^{(i)}= {\rm poly}(H,A_k)$ are generated by a quadratic-linear Hamiltonian $H$ and set of Lindbladians $A_k$, 
we can always find a polynomially large bound for the norm $\|D_\alpha^{(i)}(t')\| \leq {\rm poly} (L)$. Thus, there always exists a discretization step 
\be
\Delta\tau \leq \frac{\epsilon}{t\ {\rm poly}(L)}
\ee
that keeps the error in Eq.~\eqref{eq:error_bound_final} arbitrarily small, suppressed at least polynomially with the number of modes $L$.

Let us now estimate the amount of computational resources required to perform the above sampling procedure. For each sample, the algorithm randomly chooses a unitary trajectory according to the given prescription for each class EC1--EC3 and, according to the Terhal-DiVincenzo algorithm, produces outputs from the free-fermion distribution in Eq.~\eqref{eqm:unitary_sampling_probability}.
In particular, it samples the output at site $i$ conditioned upon the outcomes sampled at sites $j < i$, for which the marginal probabilities should also be computed. Consider cases of EC1 and EC2 that do not require ancillas.
Once the matrix $T$ is obtained
in Eq.~\eqref{eq:c_transform}, the number of steps to compute the distribution is equal to $C' = L \times O(L^3) = O(L^4)$, where the factor $O(L^3)$ is the upper bound on the time it takes to compute a Pfaffian of an $O(L)\times O(L)$ matrix.
Further, the runtime for obtaining the matrix $T$ is proportional to $t/\Delta \tau \times M(2L)$, where $M(n)\lesssim O(n^3)$ is the time for $n\times n$ matrix multiplication.
In sum, the total bound on the runtime for each trajectory is bounded as
$
C \sim O(L^4) + O\left(L^3t/{\Delta \tau}\right).
$
Choosing $\Delta \tau = \epsilon/(t\times \mathrm{poly}(L))$, the runtime is
\begin{align}\label{eq: poly_bound}
    C\leq  \mathrm{poly}\left( L,  \frac{t^2}{\epsilon} \right).
\end{align}
For EC3, the derivation is the same up to adjusting the system size to include the ancilla modes, $L\to L+t/\Delta\tau$. This case also has a similar polynomial upper bound on the classical runtime in the form of Eq.~\eqref{eq: poly_bound} as long as the evolution time $t$ is polynomial.

Finally, let us analyze the case when the conditions of classes 1--3 are violated. Strictly speaking, then
the stochastic method fails as it generally maps the problem to a nonquadratic fermionic evolution, which is not believed to be simulable for arbitrarily long time using only polynomial overhead in the number of fermion modes and computational time. . 
However, we can still efficiently simulate the system after this mapping if the product of evolution time $t$ and the correction rate $\delta\Gamma$ (of processes violating easiness conditions) remains small.
In particular, if the product is bounded as $\delta\Gamma t < c/L^2$ for some constant $c$, the dynamics remains classically easy.
To obtain this result, we consider a more general form of stochastic unraveling in Eq.~\eqref{eq:stoch_unitary_map} with nonunitary unraveling.
This formula can be Taylor expanded as $\mathcal V_{\rm st} \to \mathcal V'_{\rm st} = \mathcal V_{\rm st}+\delta\Gamma \tau K_1+(\delta\Gamma \tau)^2 K_2+\dots$, where $K_n$ are local correction superoperators and $\tau$ is the evolution time.
Therefore, we can update the bound for the operator norms in Eq.~\eqref{eq:bound_on_D} as $\|D^{(i)}_\alpha(t,t')\| \leq \|D^{(i)}_\alpha(t')\|+\delta\Gamma \tau\|K_1 D^{(i)}_\alpha(t')\|+(\delta\Gamma \tau)^2\|K_2 D^{(i)}_\alpha(t')\|+\dots$.
Since the operator $D^{(i)}_\alpha$ involves at most $k_{m}$ fermion Fock operators and the action of $K_n$ involves at most four fermion operators, we see that $\|K_n D^{(i)}_\alpha(t)\|\leq O(L^{k_{ m}+2n})$.
Therefore, the norm $\|D^{(i)}_\alpha(t,t')\|$ is always bounded by a ${\rm poly}(L)$ value if $\delta\Gamma < c/(L^2\tau)$, where $\tau = t-t'\leq t$.
This result leads to Eq.~\eqref{eq: poly_bound}.
As a result, if the dissipation is close enough to the symmetric point, the evolution remains classically easy.
This result may be helpful for analyzing the precision needed for implementing this dynamics in intermediate-scale quantum devices.

\section{Hard class \label{sec:hard}}
 We have so far demonstrated cases when the probability distribution generated by the Lindblad equation is efficiently simulable on a classical computer. Can we extend these proofs to the most general case of quadratic $A_k$'s? Since quadratic operators $A_k$ correspond to single-fermion jumps in many cases, one may expect that the problem can be solved in the single-particle sector, similar to unitary free-fermion dynamics. However, such an intuition is incomplete. A simple explanation can be obtained using the Fermi exclusion principle that requires the transition between two modes to depend on the occupation of the target mode; thus a quadratic Lindbladian jump operator can induce many-body correlations in the system that quickly become classically intractable.

\subsection{Reduction to a generic quantum circuit} 

We now provide a rigorous argument for worst-case hardness based on the equivalence of dynamics under classes $H(t)$ and $A_k(t)$ in Eq.\,\eqref{eq:Lindblad_equation} on the one hand and universal quantum computing on the other. Let us start with the simplest map utilizing quadratic Hamiltonians. We distribute all modes into $L/2$ pairs, each pair corresponding to a logical qubit in the state $|0\>_L = |01\>$ or $|1\>_L = |10\>$. Then, utilizing only quadratic Hamiltonians and Lindblad jump operators, we can implement any quantum circuit with arbitrary precision. 
Thus, by showing the equivalence of the dynamics to universal quantum computation, we obtain hardness results for both estimating time-evolved local observables $\langle O(t) \rangle$ and sampling from the time-evolved state in any local basis.
The obtained hardness result is therefore on par with the best complexity-theoretic evidence that simulating quantum circuits (in both senses) is hard.

First, using single-fermion hopping between the two sites of a qubit, we can reproduce arbitrary single-qubit operations \cite{Underwood2012}. Second, to approximately generate a desired two-qubit gate, we can use hopping combined with a quadratic Lindblad operator. In particular, assigning the two-qubit logical states $|00\>_L = |0101\>$, $|01\>_L = |0110\>$, $|10\>_L = |1001\>$, and $|11\>_L = |1010\>$, the control-$Z$ gate can be implemented by simultaneously applying the hopping Hamiltonian $H = J(c^\dag_2c^{}_3+c^\dag_3c^{}_2)$ and pair-loss operator $A = \Gamma c_3c_4$ for time $t = \pi/J$, in the limit $\Gamma\gg J$. This type of dynamics can be analyzed as follows. The logical states $|01\>_L$ and $|10\>_L$ remain invariant in the course of the evolution. At the same time, in the limit $\gamma \equiv \Gamma/J \to\infty$, due to the quantum Zeno effect, the Lindblad operator's action disallows any coherent transition involving states where qubits 3 and 4 are both occupied (i.e.\ $|\cdot\cdot11\>$). As a result, the logical state $|00\>_L$ is unaffected by the evolution. Therefore, the only evolving logical state is $|11\>_L$, which acquires a phase factor $\exp(i\pi)=-1$ after time $t = \pi/J$. As a result, the effective transformation on the two logical qubits is the control-$Z$ gate
\be\label{eq:cz_gate}
|\psi\>\to U_{\pi}|\psi\>, \qquad 
U_{\pi} = \left(\begin{matrix}
1 & 0 &0 & 0\\
0 & 1 &0 & 0\\
0 & 0 &1 & 0\\
0 & 0 &0 & -1
\end{matrix}\right).
\ee
Together with arbitrary single-qubit operations, the control-$Z$ gate is enough to obtain dynamics universal for quantum computing and hence hard to approximately sample from, assuming standard conjectures in complexity theory \cite{aaronson2011computational,Bouland2019}.

Importantly, the performance of the dissipative gate relies on the Zeno-effect blockade effective for $\gamma\to \infty$. In the limit of large but finite $\gamma$, the two-qubit system has the probability $\epsilon = 2\pi/\gamma+O(\gamma^{-2})$ of ending up in states $|0011\>$ or $|0000\>$, which could result in an error in the gate (see Appendix \ref{app:gates}). To avoid computational error, we can choose the ratio $\gamma$ to be arbitrarily large by taking vanishing $J\to{\rm poly}^{-1}(L)$ for any given $\Gamma>0$. Therefore, we can keep the error below any given threshold at the cost of increased overall computation time, which remains polynomial in system size.

The proposed architecture is not unique and allows for modified and generalized realizations of logical qubits and gates. For example, if the pair decay is always present on any two neighboring modes, one may introduce an empty ancilla mode between two logical qubits in order to ensure that logical states do not decay. As another example, if the control Hamiltonian is linear in terms of Majorana operators, a logical qubit can be encoded using just a single mode. Moreover, for a reader focused on applications, we discuss below a practical modification of qubit encoding implementable in cold atoms.

{ Now let us show that pair loss is not the unique dissipation present in the hard class. In fact, this class also includes any quadratic dissipation connected to pair loss by a time-dependent linear Bogoliubov transformation, $A'(t)=\Gamma Y(t)c_1c_2Y^\dag(t)$, where $Y(t) = \exp(-iG(t))$ is a free-fermion unitary transformation and $G(t)$ is a Hermitian operator from the quadratic-linear class in Eq.~\eqref{eq: hamiltonian}. %consider an arbitrary free-fermion unitary transformation $Y(t) = \exp(-iG(t))$, where $G(t)$ belongs to the quadratic-linear class in Eq.~\eqref{eq: hamiltonian}.
To demonstrate this equivalence, we consider the pair-loss scheme described above but simultaneously replace all pair losses $A$ with $A'(t)$, the Hamiltonian $H(t)$ with $H'(t) = Y^\dag(t) H(t)Y(t)$, and instead of the initial and final states, choose states transformed by $Y(0)^\dag$ and $Y(t)$, respectively.
The resulting process has the same probability distribution; thus its complexity would be the same.
As a result, Lindbladians such as incoherent transitions $A = \Gamma c^\dag_1c_2$ or pair gains $A = \Gamma c^\dag_1c^\dag_2$ are also classically hard in combination with free-fermion dynamics (see Table \ref{tab:summary}).}

\subsection{Robustness of the hardness result} 

\begin{figure}[]
    \centering
    \includegraphics[width=0.4\textwidth]{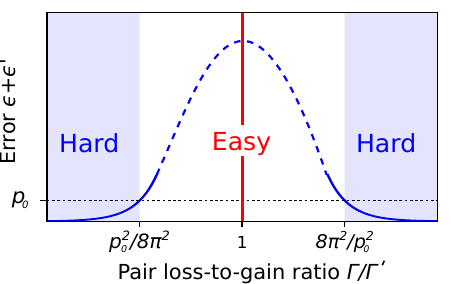}
    \caption{\textbf{Complexity phase diagram for a fermionic system with simultaneous pair losses and gain}. The plot illustrates the connection between complexity of simulation and the hypothetical dissipative control-$Z$ gate error $\epsilon+\epsilon'$ (both axes use log scale). When the error is smaller than the best-known two-qubit error-correction threshold $p_0$, the worst-case system dynamics is equivalent to that of a fault-tolerant quantum computer (blue shaded regions) and, according to existing complexity conjectures, is classically computationally hard to simulate. In contrast, when the rates of gain and loss are exactly equal, the problem belongs to EC1 (vertical red line) with the effective classical algorithm provided in the text. The result for the unshaded region remains inconclusive. The dashed line represents qualitative extrapolation.}
    \label{fig2}
\end{figure}

The error associated with imperfect Zeno blockade cannot be arbitrarily suppressed by slowing down the computation if there are small generic corrections to the dissipative dynamics.
These corrections can be viewed as the presence of additional Lindblad jump operators with total rate $\Gamma'$.
Such terms generate additional transitions with the probability $\epsilon' \sim \pi\Gamma'/J$, where $\Gamma'$ is the combined rate of added operators $A'$ and/or other errors. In contrast to the imperfect-Zeno-blockade error, this type of error diverges for small $J$. Therefore, there is an optimal value $J\sim \sqrt{\Gamma'/\Gamma}$ that minimizes the overall gate error to $\epsilon+\epsilon'\sim O(1)$, including, besides standard errors, \textit{leakage} into states outside of the logical Hilbert space. 
For fixed $\Gamma$, there always exists a choice of $\Gamma'\sim O(1)$ that keeps the error below any provided threshold, $\epsilon+\epsilon' <p_0$, where $p_0>0$. According to the leakage threshold theorem in Ref.\,\cite{aliferis2005fault}, which is a generalization of earlier standard threshold results \cite{aharonov1996fault,knill1998resilient,terhal2015quantum}, a universal set of such gates can be used to implement fault-tolerant quantum computing.
Therefore, there are instances of Lindblad evolutions that remain hard to simulate for arbitrarily long times. 

\begin{figure*}[t]
    \centering
    \includegraphics[width=1\textwidth]{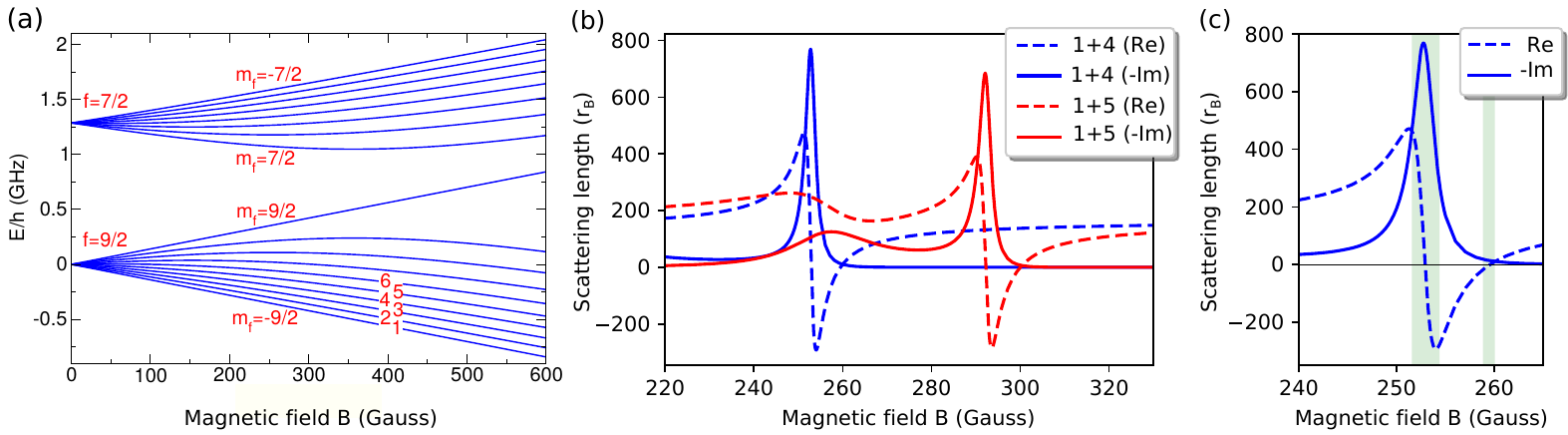}
    \caption{\textbf{Magnetic Feshbach resonance for $^{40}$K atoms.} (a) The hyperfine-Zeeman energy levels $E/h$ (GHz) of the $f=$ $9/2$ and $7/2$ manifolds versus magnetic field $B$ (Gauss) \footnote{note also that we use G (Gauss) as the magnetic field unit in this paper because of its near-universal usage among groups working in this field, 1 G=$10^{-4}$ T.}, labeled 1, 2, $\ldots$ in order of increasing energy. The levels labeled $1,2,\ldots$ have spin projections $m_f=-9/2,-7/2,\dots$, respectively.  The spin and Zeeman coupling parameters are taken from Ref.\,\cite{Arimondo1977}. (b) Scattering lengths of two $^{40}$K atoms. The solid and dashed lines represent  $-\Im (\tilde{a}_0)\propto K_2$ and $\Re (\tilde{a}_0)$, respectively; blue and red represent the $1+4$ and $1+5$ channels, respectively. (c) Magnetic Feshbach resonance for the $1+4$ channel. The shaded regions depict magnetic fields where the elastic interaction between two atoms is smaller than the pair-loss rate, marking the regime of dissipative fermionic dynamics.
    All lengths are provided in Bohr-radius units $r_{\rm B} =5.29\times 10^{-9}$ cm, and the collision energy is $E/k_\mathrm{B}=1\mu$K, where $k_\mathrm{B}$ is the Boltzmann constant.}  
   \label{fig3}
\end{figure*}

One particular example of a dissipative correction to ideal dynamics is the presence of pair gain $A'_{ij}=\Gamma'c^\dag_ic^\dag_j$ that acts on exactly the same sites as pair loss $A_{ij}$. In this case, the minimum error is $\epsilon+\epsilon' = \sqrt{8\pi^2\Gamma'/\Gamma}$ and the problem remains hard for a classical computer if $\Gamma' \leq p^2_0/8\pi^2 \Gamma$. Since the entangling gate is also implementable using pair gain instead of loss, this inequality also works after replacing $\Gamma$ by $\Gamma'$. Thus, the problem of simulating the evolution in the regions $\Gamma'/\Gamma\leq p^2_0/8\pi^2$ and $\Gamma'/\Gamma\geq 8\pi^2/p^2_0$ is classically hard. The complexity for the rest of parameter space remains an open problem. Notably, there exists at least one point in this range, $\Gamma=\Gamma'$, that is easily simulable by a classical computer since it is in EC1. Therefore, by changing the ratio $\Gamma'/\Gamma$, we can potentially induce a complexity phase transition. Figure \ref{fig2} illustrates the connection between gate error and sampling complexity.

{ Summarizing, we established quantum computational universality of quadratic dissipation combined with free-fermion dynamics, where dissipation replaces the unitary interactions between fermions. This result opens a possibility of using simple dissipation processes as a resource for quantum computing. In the following section, we illustrate the feasibility of this proposal by considering a system of cold atoms.}

\section{Application to cold atoms \label{sec:atoms}}

{ In this section, we discuss an experimentally relevant system where naturally occurring dissipation can be a source of computational hardness. In particular, we study trapped cold fermionic atoms and consider nonunitary pair collisions as a viable dissipation mechanism. Collision dynamics can be controlled by a magnetic Feshbach resonance; tuning into the resonance can suppress unitary interactions and amplify the loss rate, thus physically implementing the Zeno regime we studied in the previous section. Therefore, we show that cold atoms harbor a natural way of implementing quantum computing using dissipation-assisted operations. While one may combine dissipation with elastic interactions to increase the fidelity of the gates, our complexity analysis shows that dissipation alone is sufficient. Below we provide details on how to implement collision control and to construct the gates. We also estimate the resulting gate  error for a particular system.}

\subsection{Feshbach resonance \label{sec:feshbach}}

The Feshbach resonance provides a perfect tool to manipulate interactions between trapped atoms. Several mechanisms are available for practical use including magnetic, optical, and orbital Feshbach resonance  \cite{chin2010feshbach,Giorgini2008,cappellini2019coherent}. For concreteness, we study only magnetic resonance here. The other two mechanisms have a qualitatively similar effect on atomic interactions. We study magnetic Feshbach resonance since it does not involve laser transitions and potentially has smaller scope for error.

We also require a fermionic atom that can be cooled, trapped and prepared in specific spin states with the requisite interaction properties.  A promising example we illustrate here is the $^{40}$K atom in its $^2$S atomic ground state, which has an electron spin $S=1/2$ and nuclear spin $I=4$, giving rise to total spin $f=$ $9/2$ or $7/2$.  The Zeeman substructure of the ground-state hyperfine manifold, shown in Fig.~\ref{fig3}(a), gives rise to magnetically tunable Feshbach resonances in the interaction of two atoms for controlling elastic and dissipative collisions \cite{chin2010feshbach,Giorgini2008}. This example serves as a proper illustration of how dissipation and interaction rates can be tuned, and alternative schemes for both alkali and alkaline-earth (like) atoms can be proposed using other types of Feshbach resonance.

It is straightforward to set up and numerically solve for the scattering and bound states of two $^{40}$K atoms, including the atomic electrons and nuclear spins, their mutual interactions, and the mass-scaled adiabatic Born-Oppenheimer molecular potentials for the $^1\Sigma_g^+$ and $^3\Sigma_u^+$ states \cite{Falke2008}.  We use the standard coupled channels method \cite{Stoof1988,chin2010feshbach} to set up the full spin Hamiltonian and solve the matrix Schr{\"o}dinger equation for the scattering states.  Such models, when calibrated against bound state and scattering data, provide highly accurate predictions of the properties associated with magnetically tunable Feshbach resonance states used to tune the scattering properties of two ultracold K atoms \cite{Loftus2002,Regal2003a,Derrico2007,Falke2008,Ludewig_PhD_2012,Krauser2017,Tanzi2018,Chapurin2019}. 

The collision of two $^{40}$K atoms is characterized by the quantum numbers of the two separated atoms with resultant total spin projection $m_F = m_{f1}+m_{f2}$ and relative angular momentum, or ``partial wave,"  $\ell$ and $m_\ell$.  States with the same total angular momentum projection $M_{tot} = m_F + m_\ell$ are coupled in the molecular Hamiltonian for two atoms.  Two like fermions collide with odd partial waves, e.g., the $p$ wave with $\ell=1$, whereas two unlike fermions can collide via even or odd partial waves, including the $s$ wave with $\ell=0$.  A collision ``channel'' is defined by the partial wave and the spin quantum numbers of the two atoms. Strong pair loss via spin-exchange interactions is only possible if there is a product channel available with the same $M_{tot}$ and $\ell$ as the entrance channel; otherwise weak spin-dipolar spin relaxation is possible where $m_F$ and $m_\ell$ change by two units, conserving $M_{tot}$.   Strong $s$-wave spin-exchange relaxation is possible for the spin channels $1+4$ or $1+5$, but not for $1+2$ or $1+3$ channels; furthermore, only weak $s$-wave spin-dipolar spin relaxation is possible in the $1+3$ channel. We also do not consider the much weaker $p$-wave spin relaxation in these channels at ultracold temperatures (see below).   Consequently, we now concentrate on the $1+4$ and $1+5$ $s$-wave collisions for engineering dissipative collisions with weak on-site unitary interaction.

Very-low-energy $s$-wave elastic and dissipative collisions in the threshold regime are adequately described by a complex scattering length $\tilde{a}_0$ \cite{Balakrishnan1997,Bohn1997}, defined as the $k \to 0 $ limit of the energy-dependent complex scattering length \cite{Hutson2007,Idziaszek2010,Frye2015,Nicholson2015},
\be\label{eq:scattlength}
\tilde{a}_k = \frac{1}{ik} \frac{1-S(k)}{1+S(k)} \,.
\ee
Here $\hbar k$ is the relative collisional momentum for a collision of two atoms with reduced mass $\mu$ at energy $E=\hbar^2 k^2/(2\mu)$, and $S(k)$ is the diagonal element of the unitary $S$-matrix for the collision channel in question. In this subsection, we keep track of $\hbar$ for added clarity. The coupling constant for  the low-energy zero-range regularized pseudopotential approximation for atomic interactions is $g=2\pi\hbar^2 \mathrm{Re}(\tilde{a}_0)/\mu$ \cite{Huang1957,chin2010feshbach,Giorgini2008}. The dissipative loss rate $\dot{n}_1=\dot{n}_2=-K_2 n_1 n_2$ from colliding atoms in a gas with densities $n_1$ and $n_2$ is given by the rate constant $K_2 = -4\pi\hbar \Im(\tilde{a}_0)/\mu$ \cite{Hutson2007,Idziaszek2010} (since $\mathrm{Im}(\tilde{a}_0)$ is zero or negative, $K_2$ is positive definite; $g$ can be positive or negative).

Using counterpropagating laser beams, it is possible to construct an array of trapping cells in an optical lattice structure \cite{Bloch2005}.  Each cell is approximately harmonic and, in its ground state, may hold exactly zero, one, or two atoms.  The scattering length formulation can readily be adapted to two atoms in an optical lattice cell to calculate the interaction energy or dissipative loss rate.  For a harmonic trap with frequency $\nu=\omega/(2\pi)$, the analytic interaction energy for the lattice ground state from the zero-range pseudopotential is $\left ( 3/2 + (2/\sqrt{\pi} \right )\mathrm{Re}(\tilde{a}_0)/d)\hbar\omega$, where the harmonic length $d=\sqrt{\hbar/(\mu\omega)}$ \cite{Busch1998}.  If the lattice zero-point energy $3\hbar\omega/2$ is large enough, $\mathrm{Re}(\tilde{a}_k)$ may need to be evaluated at the lattice eigenenergy instead of taking $\mathrm{Re}(\tilde{a}_0)$ in the $k \to 0$ limit \cite{Blume2002,Bolda2002}.     The decay rate $\Gamma$ of an atom from the cell is given by $K_2 \bar{n}$, where $\bar{n} = \int d\mathbf{r} |\Psi_0(\mathbf{r})|^4 = 1/(\pi^{3/2}d^3)$ can be interpreted as a mean local density in the ground state of the lattice cell with wave function $\Psi_0(\mathbf{r})$ \cite{Tiesinga2000}. 

The figure of merit for our dissipative quantum gate, the opposite requirement from that of Ref.\,\cite{Tiesinga2000}, is that $|\mathrm{Im}(\tilde{a}_0)/\mathrm{Re}(\tilde{a}_0)| \gg 1$.   This is possible to achieve using two $^{40}$K atoms in states 1 and 4 or states 1 and 5, as we now show from our coupled channel calculations.  Using the mass scaled potentials of Ref.\,\cite{Falke2008} and including $s$  and $d$ waves ($\ell=$ 0 and 2) in the coupled channels expansion for unlike spin species gives the scattering lengths shown in Fig.~\ref{fig3}(b).

There are two regimes where the interaction energy proportional to $\Re(\tilde{a}_0)$ is small and the dissipation rate proportional to $\Im(\tilde{a}_0)$ is large.  These are in the ``core'' of the resonance, rounded into a dispersive shape by the decay, and near the zero crossing where $\Re(\tilde{a}_0)=0$.  

In order to get a sense of time scales, we can assume a harmonic length on the order of 100 nm, for which $\bar{n}\approx 2\times 10^{14}$ cm$^{-3}$.  If we take the van der Waals length $R_\mathrm{vdW}$ of two $^{40}$K atoms, 3.4nm \cite{chin2010feshbach}, as a ``typical'' size for $\Im(\tilde{a}_0)$, then $K_2 \approx 1.3 \times 10^{-10}$ cm$^3/s$, giving a decay time of $\Gamma^{-1} =$ 40 $\mu$s.  The next subsection discusses how such magnitudes could enable the realization of dissipation-assisted quantum computing.  

We note that there are other spin channels for $^{40}$K and in other species where Feshbach tuning of a favorable ratio $\mathrm{Im}(\tilde{a}_k/d)/\mathrm{Re}(\tilde{a}_k/d)\gg 1$ could be feasible.  This may be possible for like fermions, where only $p$-wave channels are available.  However,  $p$-wave interactions,  treated by Eq.~\eqref{eq:scattlength} with a $p$-wave $S$-matrix element, are typically suppressed by a factor on the order of $k^2 R^2_\mathrm{vdW}$ relative to the range of $s$-wave processes, due to the threshold law for $p$ waves \cite{DeMarco1999,Regal2003b,Idziaszek2006,Idziaszek2010}.  This suppression factor is on the order of 0.001 for $^{40}$K atoms with an energy on the order of 1 $\mu$K, so it would be harder to find ranges suitable for experimental control.

 \begin{figure}[t!]
    \centering
    \includegraphics[width=0.45\textwidth]{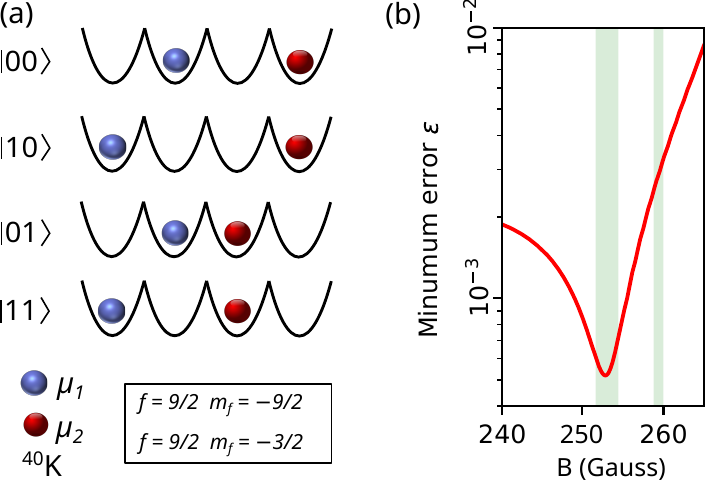}
    \caption{ \textbf{Dissipative gate utilizing cold atoms.} (a) Two-qubit logical states encoded using a pair of atoms in distinct states occupying four neighboring sites. (b) The upper bound for the total dissipative gate error as a function of magnetic field for $^{40}K$, given the background error rate $\Gamma' = 10^{-2}$s$^{-1}$. The shaded areas describe the weak interactions regime shown in Fig.~\ref{fig3}(c).}
    \label{fig4}
\end{figure}

\subsection{Dissipation-assisted quantum computing}

We now modify the idea from Sec.\ \ref{sec:hard} to make it feasible for cold atomic systems. 
We limit our attention to two distinct states of trapped atoms, denoted here as $\mu_1$ and $\mu_2$. A single trap is described using the following four basis states: the empty state $|0\>_i$, single-occupied states $|\mu_\alpha\>_i = c^\dag_{i\mu_\alpha}|0\>_i$, $\alpha=1,2$, and the double-occupied state $|\mu_1\mu_2\>_i=-|\mu_2\mu_1\>_i = c^\dag_{i\mu_1}c^\dag_{i\mu_2}|0\>_i$. We consider the lattice Hamiltonian $H = H_{\mu_1}+H_{\mu_2}+V$, where $H_{\mu} = \sum_{\<ij\>}J^\mu_{ij}(t) (c^\dag_{i\mu}c^{}_{j\mu}+{\rm h.c.})+ \sum_i \Delta^\mu_i(t)c^\dag_{i\mu}c^{}_{i\mu}$. The quantities $J^\mu_{ij}(t)$ are real tunneling amplitudes, and $\Delta^\mu_i(t)$ are on-site potentials, both of which can depend on the atomic state $\mu$. 
Two distinct atoms located in the same trap are subject to elastic interactions $V = E\sum_in_{i\mu_1}n_{i\mu_2}$, where $E$ is the interaction energy and $n_{i\mu} = c^\dag_{i\mu}c^{}_{i\mu}$ is the $\mu$-occupation number of the $i$th trap.
As shown in Sec.\ \ref{sec:feshbach}, the interaction $E$ may be made to vanish for a specific pair of states $\mu_1$ and $\mu_2$ by, for example, manipulating the magnetic field as shown in Fig.~\ref{fig3}(c). Also, atoms in the same trap undergo pair loss with rate $\Gamma$, described by Lindblad jump operators $A_i = \Gamma c_{i\mu_1}c_{i\mu_2}$. For $E=0$, the entire dynamics is described by the master equation Eq.~\eqref{eq:Lindblad_equation}.

The computational scheme utilizes pairs of sites to encode individual logical qubits. The logical qubit states are $|0\>_L = |0\>|\mu_j\>$ and $|1\>_L = |\mu_j\>|0\>$, irrespective of the atom's type $\mu_j$. 
Single-qubit gates can be performed using the local potentials and coherent hopping between logical qubit sites.

Previously, we discussed how to use Feschbach resonance to induce and control the dissipation of alkali atoms using the example of $^{40}$K. We now step aside to describe how a generic entangling gate works for a larger group of atoms, assuming that the same level of control can be applied to them as well.
We consider two distinct ways of constructing entangling gates, depending on the atomic electronic structure.
The first method we consider is designed for alkaline-earth(like) atoms such as $^{87}$Sr \cite{targat2006accurate} and $^{173}$Yb \cite{Scazza2014observation,riegger2018localized}.
We can use nuclear-spin polarized metastable states $^1S_0$ and $^3P_0$ as the two species $\mu_1$ and $\mu_2$ \cite{daley2008quantum} in order to apply a species-dependent hopping term $J^\mu_{ij}$. The scattering length between $\mu_1$ and $\mu_2$ can then be potentially tuned by optical or orbital Feshbach resonances to be purely imaginary. Alternatively, one could use $^3P_2$ instead of $^3P_0$, in which case a magnetic Feshbach resonance is also an option.
However, this method has limited applicability to alkali atoms, for example $^{87}$Rb \cite{anderlini2007controlled} or $^{40}$K (described above), where the states $\mu_i$ are encoded into different angular momentum projections $m_f$.
This is because a state-dependent lattice in alkali atoms \cite{lee2007sublattice} can exhibit significant single-atom dissipation rates.
For alkali atoms, therefore, we propose an additional scheme that does not rely on an internal-state-dependent lattice and uses the same lattice potential for both states.
In order to implement entangling gates, we make use of single-qubit rotations with single-site resolution. This can be achieved using two-photon Raman transitions induced by focused laser beams or other similar techniques \cite{yavuz06, zhang2006manipulation, gorshkov2008coherent}. Both schemes can be used interchangeably. 

We now give details of the two schemes. Consider the four two-qubit logical states $|00\>_L = |0,\mu_1,0,\mu_2\>$, $|01\>_L = |0,\mu_1,\mu_2,0\>$, $|10\>_L = |\mu_1,0,0,\mu_2\>$, and $|11\>_L = |\mu_1,0,\mu_2,0\>$ [see Fig.~\ref{fig4}(a)], where the comma separates states of individual traps. For the first scheme, an entangling control-$Z$ gate is performed in a single step by applying the hopping $H = J(c^\dag_{2\mu_2}c^{}_{3\mu_2}+{\rm h.c.})$ for state $\mu_2$ between traps 2 and 3 for time $t = \pi/J$.
As a result, the states $|00\>_L$ and $|01\>_L$ remain invariant under the evolution, while any transitions involving the state $|01\>_L$ are blocked by the quantum Zeno effect. In the limit $\Gamma/J \to\infty$, the overall unitary operation in the logical Hilbert space is described by the control-Z gate in Eq.~\eqref{eq:cz_gate}. 
For the second scheme, the control-Z gate can be applied in three steps: (1) apply the state-independent hopping $H = J\sum_i(c^\dag_{2\mu_i}c^{}_{3\mu_i}+{\rm h.c.})$ between traps 2 and 3 for time $t = \pi/(2J)$; (2) apply a single-qubit phase gate $|\mu_1\>_3\to e^{i\pi}|\mu_1\>_3$, $|\mu_2\>_3\to |\mu_2\>_3$ on site 3; (3) repeat the first step. As a result, states $|01\>_L$ and $|10\>_L$ remain stationary, state $|00\>_L$ acquires a total phase $2\pi$ (from the phase gate and hopping), and state $|11\>_L$ acquires phase $\pi$. Thus, the second scheme also implements a control-$Z$ gate.

The performance of the gate can be disrupted by errors, including imperfect Zeno-effect error, the single-qubit phase gate error (for the second scheme), and background dissipation error. The background dissipation error can be bounded above by $\Gamma't$, where $\Gamma'$ is the background dissipation rate, and $t=\pi/J$ is the gate performance time (neglecting the time taken for the single-qubit phase gate). The single-qubit phase gate error $\epsilon_0$ is fundamentally limited by light scattering loss during the Raman transition, which depends on the characteristic linewidth $\gamma$ of the excited levels and the detuning limited by fine structure splitting $\Delta$. This error is estimated to be $\epsilon_0\sim \gamma/\Delta$. 
Deviations from perfect single-site addressability during the single-qubit phase gate can also give rise to errors, which can nevertheless be greatly reduced by subwavelength addressability techniques \cite{stokes91,schrader04,gardner93,zhang2006manipulation,lee2007sublattice, yavuz07,gorshkov2008coherent}. Finally, the error caused by an imperfect Zeno effect can be approximated by the first term in the Taylor expansion of the infidelity in $(\Gamma t)^{-1}$ (see Appendix \ref{app:gates}), leading to the total error
\be\label{eq:zeno_error}
\begin{split}
\epsilon = \epsilon_0+\Gamma't+\frac{8\pi^2}{\Gamma t}\frac 1{1+4\zeta^2}+O\Bigl(\frac{1}{\Gamma^2t^2}\Bigl),
\end{split}
\ee
where $\zeta = E/\Gamma$ is the loss-to-interaction ratio. The error can be minimized by making the choice $t = 2\pi/\sqrt{\Gamma\Gamma'(1+4\zeta^2)}$, leading to the expression
\be\label{eq:min_error}
\epsilon = \epsilon_0+4\pi\sqrt{\frac{\Gamma'}{\Gamma(1+4\zeta^2)}}.
\ee
The dependence of the second term on the magnetic field is illustrated in Fig.~\ref{fig4}(b) under the same choice of parameters as in Fig.~\ref{fig3}. For a suitable choice of magnetic field strength, the theoretical upper bound for the error can be as low as $\epsilon \sim 5\times 10^{-4}$ for the background dissipation rate $\Gamma'=10^{-2}s^{-1}$. For $^{40}$K atoms, the optical transition error  $\epsilon_0$ can be estimated using the values $\gamma \simeq 2\pi\times 6.0$ MHz and $\Delta \simeq 2\pi\times1.7$ THz \cite{tiecke2010properties}, leading to the upper bound $\epsilon_0\sim 10^{-6}$, which is an insignificant contribution to the overall error. As a result, the theoretical bound for the gate error approaches the characteristic thresholds given by many error-correcting schemes. 

{ From the comparison of elastic and inelastic rates in Fig.~\ref{fig3}(c), one can see the advantage of dissipative schemes over unitary ones. While the gate time for both elastic and inelastic schemes is proportional to the corresponding inverse collision rate, the resonant dissipation rate can be significantly larger than the accessible elastic rates. Therefore, dissipation-assisted gates can be faster and can thus experience smaller level of errors due to the background noise.
   
Summarizing, we have established that naturally occurring pair-loss processes in cold atoms can be enhanced and used as a resource for quantum computing.
This conclusion opens a possibility to improve quantum computing with cold atoms \cite{daley2008quantum} or, in absence of sufficient control, use them as a platform for quantum supremacy experiments.}

\section{Discussion}

In this work, we have demonstrated how simple forms of dissipation affect the complexity of simulation of noninteracting fermions. In particular, focusing on linear-quadratic Lindblad jump operators, we have shown the existence of two complementary complexity classes of Lindblad jump operators, easy and hard for simulation on a classical computer. Using the error-correction formalism, we showed that the hard class has a finite volume in the parameter space and tolerates the presence of small arbitrary corrections. At the same time, the easy classes may have small measure and could become hard even as a result of arbitrarily small corrections to the master equation.

We have expanded the region of classical simulability of free-fermions in the presence of Markovian errors from single-qubit loss/gain to more general quadratic-linear Lindblad jump operators.
The algorithms we devise for EC1--EC3 based on the stochastic unraveling approach provably work in polynomial time.
This shows that a large class of dissipation processes such as dephasing or single-fermion decay can be treated with the help of efficient classical algorithms.

At the same time, more complex processes are BQP-complete, where BQP stands for the class Bounded-error Quantum Polynomial time. We show this fact by explicitly constructing an entangling gate and showing the equivalence of the problem with universal quantum computation. 
We thus place limitations on the extent to which the simulability result may be extended, since we believe quantum computation is strictly more powerful than classical computation. Our detailed analysis shows that it is within the range of experimental feasibility to implement with cold atoms a quantum computer with purely dissipative atom-atom interactions, an exciting possibility for experiments in quantum computing. 
For example, dissipative quantum systems such as alkaline-earth atoms may serve in the next generation of quantum supremacy experiments. Also, our result suggests that simulating fermion dynamics may be hard for quantum particles experiencing dissipation, for example, quasiparticles in solid-state systems. Future work can explore the hardness of simulation of electronic systems with quasiparticle dynamics approximated with quadratic-linear Lindblad jump operators that include the effects of electron-electron, electron-phonon, and electron-impurity scattering processes. Alternatively, physical systems following such dynamics with high accuracy may be a future platform for quantum-computing experiments.

It may be interesting to explore the connection of our results with the theory of matchgate (free-fermion) computations and the role played by non-Gaussianity.
Quadratic fermionic Hamiltonians and single-fermion loss give rise to Gaussian operations and are hence easily simulable \cite{bravyi2011classical}.
It is known \cite{Hebenstreit2019} that any non-Gaussian fermionic state is a resource for fermionic computation, boosting the computational power of free fermions from being classically simulable to being universal for quantum computation.
Our results suggest that quadratic-linear Lindblad jump operators are non-Gaussian in general.
Therefore, it would be interesting to quantify the amount of non-Gaussianity (or ``magic'') for the Lindblad operations we study here.

Along the same lines, one can quantify a different resource for nonclassicality, such as a suitable measure of entanglement for open fermionic systems.
Efficient sampling from the full output distribution in arbitrary bases can allow for efficient computation of certain measures of entanglement such as R\'enyi entropies \cite{Brydges2019}.
Relatedly, Ref.\,\cite{Alba2020} has studied the logarithmic negativity for free fermions with gain/loss Lindblad terms.

Further, one may also consider how the complexity of simulating dynamics under quadratic Lindbladians changes with time.
Since the system starts off in a Fock state that is easy to sample from, and dynamics under quadratic Hamiltonians with quadratic Lindblad jump operators can generate states that are hard to sample from, one can see a dynamical transition in sampling complexity \cite{PhysRevLett.121.030501,Maskara2019}.
It is worthwhile to investigate whether these transitions are sharp or coarse (as defined in Ref.\,\cite{Maskara2019}) since this can identify what ``universality class'' free fermions with noise belong to.
Techniques such as Lieb-Robinson-like bounds for the evolution of free particles with dissipation \cite{Poulin2010,Barthel2012} would be relevant here.

Another exciting direction is the study of worst-to-average-case equivalence in complexity, which seeks to understand the complexity of typical instances as opposed to worst-case instances \cite{Bouland2019,Movassagh2019,Bouland2021}.
It would be interesting to see if the Cayley path technique of Ref.\,\cite{Movassagh2019} can be adapted to argue for average-case hardness of dissipative fermionic dynamics.

Lastly, let us address the case of bosonic particles. First, in the limit of strong dissipation, our result on computational universality applies readily to bosons. This conclusion combines with the fact that the proposed Feshbach realization scheme is generally similar for bosons and fermions. The major difference arises for identical atoms, since identical bosons also feature strong and broad resonances in $s$-wave scattering, whereas only $p$-wave resonances exist for fermions due to the Pauli principle (see, for example, Ref.~\cite{ahmed2021probing}).  Thus, the strong dissipation regime is easier to achieve for identical bosons in general. However, in the case of distinct fermions, e.g.~in different spin states, there is essentially no difference between bosons and fermions as a class \footnote{We note that individual species do show a wide range of variation due to their specific molecular physics. An example of such variation is provided by $^{87}$Rb and  $^{85}$Rb.  The former has no broad s-wave Feshbach resonances whereas the latter has a very prominent one. The difference is due to the nature of the singlet and triplet scattering lengths for the two different mass isotopes: nearly identical values suppress the Feshbach resonance for $^{87}$Rb, but quite different values permit a strong Feshbach resonance for $^{85}$Rb.}.

Next, in contrast to hardness results, classical easiness results for bosons are rather distinct from fermions. For Fock-basis measurements, free bosons are already believed to be hard to simulate in the sense of sampling from the output distribution \cite{aaronson2011computational}. This hardness equally holds for some problem variations such as using Gaussian initial states with Fock measurements \cite{Hamilton2017,Kruse2019} or homodyne measurements for non-Gaussian initial states \cite{chakhmakhchyan2017boson}.
However, since free boson evolution is not believed to be computationally universal, our results imply that quadratic dissipation boosts the computational power of free bosons in these settings to that of computational universality. Thus, it leads to another type of complexity transition between two classically hard classes. This transition is an interesting object for future studies.

Notably, there exist settings where the free-boson evolution is classically easy to simulate; an example is the situation when the  system is accessed by homodyne measurements in combination with Gaussian initial states.
Moreover, in this example, adding either single-boson or even many-body decay processes preserve classical easiness of free-boson evolution, unlike the fermionic cases considered in our work.
However, it is also an open question whether or not generic dissipative processes involving boson creation can induce a complexity transition. The analysis of these processes can be done using previous works addressing the complexity of non-Gaussian bosonic states. For example, Ref.~\cite{Chabaud2017} suggests specific non-Gaussian states that can be useful as computational resources when making homodyne measurements, while Ref.~\cite{Deshpande2021}  analyzes photon loss from the point of view of computational complexity (with Fock-basis measurements).
We keep studying this problem as well as analyzing the robustness of the resulting phases for future research.

\begin{acknowledgments}
We thank Bill Fefferman, Mohammad Hafezi, Trey Porto, and Daniel Gottesman for fruitful discussions. This work was supported by DoE ASCR Accelerated Research in Quantum Computing program (award No. DE-SC0020312), U.S. Department of Energy Award No. DE-SC0019449,  NSF PFCQC program, DoE ASCR Quantum Testbed Pathfinder program (award No.\ DE-SC0019040), AFOSR, ARO MURI, ARL CDQI, and NSF PFC at JQI.
\end{acknowledgments}

\appendix

\section{Easy Class 1 \label{app:ec1}}

In this Appendix, we analyze the convergence of the average unitary stochastic evolution to the exact Lindblad dynamics in the case of easy class 1 (EC1). 
First, we set the initial time to be zero and consider the final time $t$ being an integer multiple of timestep $\Delta\tau$. 
This assumption holds without loss of generality since $\Delta \tau$ may be adjusted appropriately to capture any particular final time. 
Then the overall evolution of unitary can be written as a product
\be\label{eq:regularization_SM}
U(t) = \prod_{n=0}^{t/\Delta\tau} U_n,
\ee
where the timestep unitary $U_n$ is expressed in terms of a time-ordered exponential
\be\label{eq:ord_exp}
U_{n} = \mathcal T\exp \Bigl(-i \int_{n\Delta\tau}^{(n+1)\Delta\tau}dtH'(t)\Bigl)
\ee
generated by the stochastic Hamiltonian $H'(t)$ in Eq.~\eqref{eq:stoch_ham1},
\be
H'(t) = H(t)+\frac {R(t)}{\sqrt{\Delta\tau}}.
\ee
Here $H(t)$ is the original time-dependent Hamiltonian, and $R(t) = \sum_k\xi_{nk}A_k(t)+\xi^*_{nk}A^\dag_k(t)$ is the normalized stochastic part, where $\xi_{nk}$ are independent complex Gaussian variables defined for times $n\Delta\tau\leq t\leq (n+1)\Delta\tau$.

Let us consider the ordered exponential expansion of
the timestep unitary in Eq.~\eqref{eq:ord_exp}:
\be\label{eq:low_order_dt_expansion}
\begin{split}
U_n &= I-i\Delta\tau^{1/2}R_n-\Delta\tau\Bigl(iH_n+\frac12R^2_n\Bigl)\\
&-\Delta\tau^{3/2}\Bigl(\mathcal PH_nR_n- \frac i6 R^3_n\Bigl)\\
&-\frac 12\Delta\tau^2\Bigl(H^2_n- \frac i3\mathcal PH_nR_nR_n-\frac 1{12}R^4_n\Bigl)\\
&+O(\Delta\tau^{5/2}),
\end{split}
\ee 
where we denote the discretized value of an operator $O_n$ and permutation sum, respectively, as
\be\label{eq:notation_sm}
\begin{split}
& O_n = \frac{1}{\Delta \tau}\int_ndt O(t)\equiv \frac{1}{\Delta \tau}\int_{n\Delta\tau}^{(n+1)\Delta\tau}dt O(t),\\
&\mathcal P O_1\dots O_m = \sum_{\sigma\in S_m} O_{\sigma(1)}\dots O_{\sigma(m)}.
\end{split}
\ee
The average over the stochastic field can be taken for each timestep independently. Therefore, the effect of the timestep unitary in Eq.~\eqref{eq:low_order_dt_expansion} is
\be\label{eq:stochastic_decomposition}
\begin{split}
\mathbb E U_n\rho U^\dag_n = \Bigl(I+\mathcal L_n\Delta\tau&+\frac12 \mathcal L^2_n\Delta\tau^2\Bigl)\rho\\
&+\mathcal D_n\rho\Delta\tau^2+O(\Delta\tau^3).
\end{split}
\ee
In the equation above, $\mathcal L_n$ is the generator of the original Lindblad equation,  $\lim_{\Delta\tau\to0}\mathcal L_n = \mathcal L(n\Delta\tau)$, expressed as
\be\label{eq:linblad_EC1}
\begin{split}
\mathcal L_n\rho  = -i[H_n,\rho]&+\sum_k\Bigl(A_{kn}\rho A^\dag_{kn}-\frac 12\{A^\dag_{kn}A_{kn},\rho\}\Bigl)\\
&+\sum_k\Bigl(A^\dag_{kn}\rho A_{kn}-\frac 12\{A_{kn}A^\dag_{kn},\rho\}\Bigl),
\end{split}
\ee
and $\mathcal D_n$ represents the lowest-order correction occurring due to the timestep being nonzero:
\be\label{eq:ec1_correction_full}
\begin{split}
\mathcal D_n\rho =&\frac 14\sum_{kk'}\Bigl( A^\dag_{k'n}A^\dag_{kn}\rho [A_{k'n},A_{k}]+ A^\dag_{k'n}A_{kn}\rho [A_{k'n},A^\dag_{kn}]\\
&\qquad+A_{k'n}A^\dag_{kn}\rho [A^\dag_{k'n},A_{k}]+ A_{k'n}A_{kn}\rho [A^\dag_{k'n},A^\dag_{kn}]\Bigl)\\
& +\sum_k\Bigl({A_{kn}}\rho V_{kn}+V_{kn}\rho {A_{kn}}+V^\dag_{kn}\rho{A^\dag_{kn}} +{A^\dag_{kn}}\rho V^\dag_{kn}\Bigl)\\
& +W_n\rho+\rho W_n^\dag.
\end{split}
\ee
Here we use the notation
\be
\begin{split}
V_{kn} &= \sum_{k'}\frac 14\{A_{kn}^\dag,\{A^\dag_{k'n}A_{k'n}\}\}-\frac 16\mathcal PA^\dag_{kn}A^\dag_{k'n}A_{k'n},\\
W_n &=-\frac i6\sum_k\Bigl(\Bigl[[H_n,A_{kn}],A_{kn}^\dag\Bigl]+\{H_n,A_{kn}^\dag A_{kn}\}\Bigl)\\
&-\frac 18\Bigl(\sum_k \{A^\dag_{kn},A_{kn}\}\Bigl)^2+\frac 1{48}\sum_{kk'}\mathcal P A^\dag_{kn}A_{kn}A^\dag_{k'n}A_{k'n}.
\end{split}
\ee
The overall expression in Eq.~\eqref{eq:ec1_correction_full} can be written in a compact form,
\be\label{eqapp:correction_form}
\mathcal D_n\rho = \sum_\alpha D^{(1)}_{\alpha n}\rho D^{(2)}_{\alpha n},
\ee
where $D^{(i)}_{\alpha n} = {\rm poly}(H_n,A_{kn})$ are polynomials of degree less than four.

The averaged stochastic map in Eq.~\eqref{eq:stochastic_decomposition} can be rewritten as a continuous evolution and then decomposed using Dyson series for the small parameter $\Delta\tau$,
\be\label{eqapp:V_sp_decomposition}
\begin{split}
\mathbb E &\mathcal V_{\rm st}(t_2,t_1) = \mathcal T\exp\Bigl(\int_{t_1}^{t_2}dt'(\mathcal L(t')+\mathcal D(t')\Delta\tau)\Bigl)+O(\Delta\tau^2)\\
&= \mathcal V(t_2,t_1) + \int_{t_1}^{t_2}dt'\mathcal V(t_2,t') \mathcal D(t')\mathcal V(t',t_1)\Delta\tau + O(\Delta\tau^2),
 \end{split}
\ee
where the generators $\mathcal L(t)$ and $\mathcal D(t)$ are continuous versions of the operators in Eq.~\eqref{eq:linblad_EC1} and Eq.~\eqref{eq:ec1_correction_full}, in which the $\Delta\tau$-averaged operators $A_{kn}$ and $H_n$ are replaced by the corresponding instantaneous values at time $t$, i.e.\ $A(t)$ and $H(t)$,  respectively. To obtain the expression in Eq.~\eqref{eq:V_sp_decomposition2}, we recursively replace $\mathcal V(t_2,t')$ and $\mathcal V(t',t_1)$ on the right-hand side by their stochastic average and collect all $O(\Delta\tau^2)$ terms.

\section{Easy Class 2 \label{app:ec2}}

In this Appendix, we analyze the convergence of the average stochastic unitary evolution to the Lindblad dynamics in the case of easy class 2 (EC2). The single timestep evolution averaged over stochastic unitaries in Eq.~\eqref{eq:stochastic_unitary} is equivalent to the map
\be\label{eq:easy_class2_expect}
\begin{split}
\mathbb EU_n\rho U_n^\dag &= U_n^0\Bigl(\rho+\int_ndt\sum_k \Gamma_k(t)\Bigl(Y_{k}(t)\rho Y^\dag_{k}(t) - \rho\Bigl)\Bigl)U_n^{0\dag}\\
 & = \Bigl(I+\mathcal L_n+\frac 12\mathcal L^2_n\Bigl)\rho+\mathcal D_n\rho \Delta\tau^2+O(\Delta\tau^3),
\end{split}
\ee
where the target Liouville operator is
\be
\mathcal L_n\rho = -i[H_n,\rho]+\sum_k A_{kn}\rho A^\dag_{kn} - \Gamma_{kn}\rho.
\ee
The correction now takes the form
\be\label{eq:D_correction2}
\begin{split}
\mathcal D_n\rho=  &\sum_k\Bigl(A_{kn}\rho C_{kn}+C^\dag_{kn}\rho A_{kn}^\dag\Bigl)\\
&-\frac 12 \sum_{kk'}A_{kn} A_{k'n}\rho A^\dag_{k'n} A^\dag_{kn}-\frac 12 \Gamma_n^2\rho,
\end{split}
\ee
denoting $C_{kn} = \Gamma_n A^\dag_{kn}+\frac i2[A^\dag_{kn},H_n]$ and $\Gamma_n = \Delta\tau^{-1}\int_n dt\sum_k\Gamma_k(t)$. This expression has the form of Eq.~\eqref{eqapp:correction_form} with operators $D^{(i)}_{\alpha n}$ being a sum of products of at most four Majorana fermion operators. 

\section{Easy Class 3 \label{app:ec3}}

In this Appendix, we analyze easy class 3 (EC3) and show the convergence of the system-ancilla stochastic evolution under the Hamiltonian in Eq.~\eqref{eq:stoch_ham1} using the stochastic operators in Eq.~\eqref{eq:ec3_stoch_field} to the dissipative dynamics with linear Lindblad jump operators. Let us start from a many-body pure state of the fermions occupying $L$ modes of the system and $L_a$ ancilla modes at time $t=n\Delta\tau$, denoting it as $|\Psi_n\>$. At the $n$th timestep, the evolution acts on the system and the $n$th ancilla mode only. Thus, the state at time $t =n\Delta\tau$ is a product state of subsystem states: (1) correlated state of $L$ system modes together with the first $n$ ancilla modes and (2) the product states of the remaining $L_a-n$ ancilla modes, i.e.\
\be\label{eq:method2_initial_state}
|\Psi_{n}\> = |\phi_n\>_{L+n}\otimes |0\>_{L_a-n}.
\ee
The evolution is governed by the Hamiltonian
\be
H'(t) = H(t)\otimes I_A+\frac 1{\sqrt{\Delta\tau}}\Bigl(K(t)+K^\dag(t)\Bigl),
\ee
where the stochastic terms are
\be
K(t) = \sum_kf_{nk}A_k(t)(\gamma_{2(L+n)}+i\gamma_{2(L+n)})
\ee
at times $n\Delta\tau\leq t\leq (n+1)\Delta\tau$, and $f_{nk}$ are independent real Gaussian variables. Then, at the $(n+1)$th step, the system-ancilla state $|\Psi_n\> = U_n|\Psi_n\>$ is
\be
\begin{split}
|&\Psi_{n+1}\> = |\Psi_{n}\>-i\Delta\tau^{1/2}K_n|\phi_n\>|1\>|0\>_{L_a-n-1}\\
&-\Delta\tau\Bigl(iH_n+\frac 12K^\dag_nK_n\Bigl)|\phi_n\>|0\>_{L_a-n}\\
&-\Delta\tau^{3/2}\Bigl(\frac 12\{H_n,K_n\}- \frac i6 K_nK^\dag_nK_n\Bigl)|\phi_n\>|1\>|0\>_{L_a-n-1}\\
&-\frac 12\Delta\tau^2\Bigl(H^2_n-\frac i3\{H_n,K^\dag_nK_n\}\\
&\qquad\qquad\qquad-\frac i3R^\dag_nH_nK_n-\frac 1{12}(K^\dag_nK_n)^2\Bigl)|\phi_n\>|0\>_{L_a-n}\\
&+O(\Delta\tau^{5/2}),
\end{split}
\ee
where we used the discrete-time operator values $H_n$ and $K_n$ obtained as in Eq.~\eqref{eq:low_order_dt_expansion}.

The interpolated continuous-time evolution for the density matrix of the system can be presented in the form
\be\label{eq:interpolation_from_discrete}
\begin{split}
\frac{d}{dt}\rho& = \frac 1{\Delta\tau} \mathbb E\,\Tr_A\Bigl(|\Psi_{n+1}\>\<\Psi_{n+1}|-|\Psi_{n}\>\<\Psi_{n}|\Bigl)\Bigl|_{n=\lfloor{t/\Delta\tau}\rfloor}\\
 & = \Bigl(I+\mathcal L_n+\frac 12\mathcal L^2_n\Bigl)\rho+\mathcal D_n\rho \Delta\tau^2+O(\Delta\tau^3),
\end{split}
\ee
where $ \lfloor{x} \rfloor $ is the floor function. The target Liouville operator is
\be
\mathcal L_n\rho = -i[H_n,\rho]+\sum_k A_{kn}\rho A^\dag_{kn} - \frac 12\{A^\dag_{kn}A_{kn},\rho\}
\ee
and the correction is
\be\label{eq:ec3_correction}
\begin{split}
\mathcal D_n\rho &=  \sum_{kk'}\Bigl(\frac 14A^\dag_kA_{k'}\rho A^\dag_{k'}A_{k} -\frac 12 A_kA_{k'}\rho A^\dag_{k'}A^\dag_{k}\Bigl)\\
&+\sum_k \Bigl(A_k\rho Q_k+Q^\dag_k \rho A_k^\dag\Bigl)+M\rho+\rho M^\dag,
\end{split}
\ee
where 
\be
\begin{split}
&Q_k = \frac 1{12}\{A^\dag_{k},A^\dag_{k'}A_{k'}\}\\
&M = \frac i{6}\sum_k \Bigl( A_k^\dag HA_k -\frac{1}{2}\{H,A_k^\dag A_k\}\Bigl)\\
&\qquad-\frac 1{12}\sum_{kk'}\Bigl(A^\dag_{k'}A_{k'}A^\dag_{k}A_{k}-\frac1{2}A^\dag_{k'}A_{k}A^\dag_{k}A_{k'}\Bigl).
\end{split}
\ee
As is the case for EC1 and EC2, the correction is described by Eq.~\eqref{eqapp:correction_form} with operators $D^{(i)}_{\alpha n}$ being a sum of products of at most eight Majorana fermion operators. 

\section{Non-Gaussianity of EC1 and EC2 \label{app:non-gauss}}

In this Appendix, we show a basic example where jump operators from easy classes (EC) 1 and 2 may be non-Gaussian processes violating Wick's theorem. First, consider a 1D nearest-neighbor hopping Hamiltonian to be the same for all examples below,
\be
H = J\sum_{n=1}^{L-1} c_{n+1}c_n+{\rm h.c.},
\ee
where $L$ is the system size and $J$ is a real hopping coefficient. For simplicity, we set an equilibrium thermal state
$
\rho(0) = Z^{-1}\exp(-\beta H)
$
as the (Gaussian) initial state, where $Z$ is the partition function and $\beta$ is an inverse temperature.

As a model of quadratic jump dissipation from EC1, we consider $2L-2$ independent incoherent jump processes between adjacent sites,
\be
F^{\rm q}_{2n} = \sqrt{\Gamma_q} c^\dag_{n+1}c_n \qquad F^{\rm q}_{2n+1} = \sqrt{\Gamma_q} c^\dag_{n}c_{n+1},
\ee
where $\Gamma_q$ is the jump rate. As required by EC1, we make these rates equal for the processes going in opposite directions (hopping to the left and to the right). 
As an example from EC2, we consider $L-1$ unitary jump operators 
\be
F_n = \sqrt{\Gamma_u}\exp(-iJ_u(c^\dag_nc_{n+1}+{\rm h.c.})),
\ee
where $J_u$ is a real hopping coefficient and $\Gamma_u$ is the  corresponding dissipation rate. 
Finally, as an example from EC3, we use the $2L$ linear dissipation operators with
\be
F^{\rm s}_{2n} = \sqrt{\Gamma_s} c_n,\qquad F^{\rm s}_{2n+1} = \sqrt{\Gamma'_s} c^\dag_n,
\ee
where $\Gamma_s\neq \Gamma'_s$ are fermion decay/gain rates. 

 \begin{figure}[t!]
    \centering
    \includegraphics[width=0.45\textwidth]{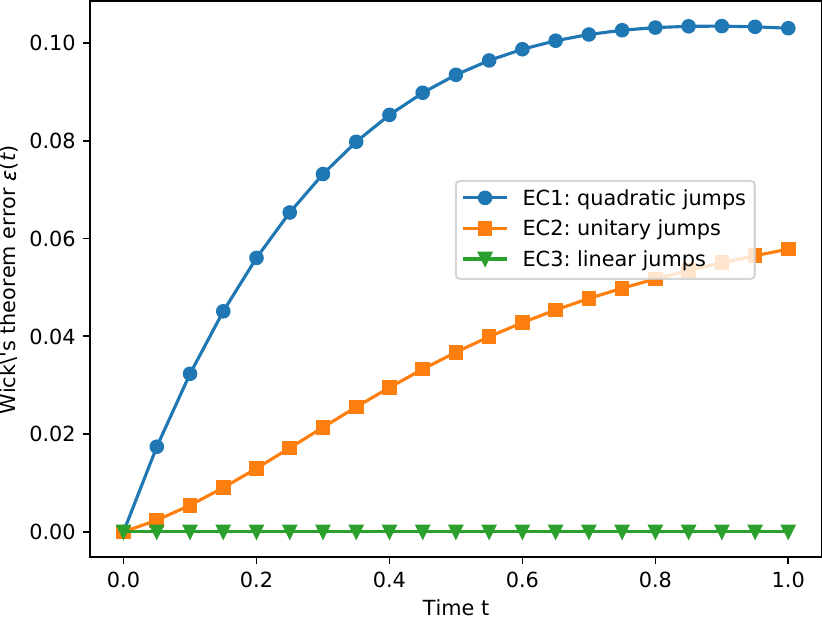}
    \caption{\textbf{Performance of Wick's theorem.} The plot shows the dependence of error in Eq.~\eqref{eq:wicks_acc} as function of time for three processes, involving quadratic jump operators from EC1 (blue, circles), unitary jump operators from EC2 (orange, squares), and linear jump operators from EC3 (green, triangles) as discussed in the text of Appendix \ref{app:non-gauss}. The systems size is $L=4$. The parameters are chosen as $\beta J = \beta\Gamma_q = \beta\Gamma_s= \beta J_u = \beta \Gamma_u = 1$, $\beta\Gamma_s' = 2$. The plot demonstrates that jump operators from EC1 and EC2 violate Wick's theorem, in contrast to the linear jump operators.}
    \label{fig1A}
\end{figure}

For all three cases, we analyze the deviation of the 4-point correlation function from the prediction given by Wick's theorem,
\be\label{eq:wicks_acc}
\begin{split}
\varepsilon(t) = \bigl|\<\gamma_1\gamma_2&\gamma_3\gamma_4\>_t-\<\gamma_1\gamma_2\>_t\<\gamma_3\gamma_4\>_t\\
&+\<\gamma_1\gamma_3\>_t\<\gamma_2\gamma_4\>_t-\<\gamma_1\gamma_4\>_t\<\gamma_2\gamma_3\>_t\bigl|,
\end{split}
\ee
where we define $\<\cdot\>_t := \Tr(\cdot \rho(t))$. The result is shown in Fig.~\ref{fig1A}. As we can see, Wick's theorem is violated for EC1 and EC2 at times $t>0$. The process from EC3 alone remains Gaussian, as discussed previously \cite{prosen2010spectral, bravyi2011classical}.
The inapplicability of Wick's theorem for EC1 and EC2 implies that the simulability results of Refs.~\cite{horstmann2013noise,vznidarivc2010exact,eisler2011crossover} do not lead to classical easiness of sampling from the full output distribution when measuring the time-evolved state in the fermion-number basis.

\section{Error analysis \label{app:error}}

In this Appendix, we first derive Eq.~\eqref{eq:erro_int_step1} and then provide the proof of the lemma in the main text.

The error can be formally expressed in terms of evolution superoperators as
\be
\begin{split}
\epsilon = \frac 12\max_{\vec r'}\sum_{\vec r}| \<\psi_{\vec r}|\mathbb E {\mathcal V}_{\rm st}(t,0)\rho_{\vec r'}(0)-\mathcal V(t,0)\rho_{\vec r'}(0)|\psi_{\vec r}\>\bigl|,
\end{split}
\ee
where $\mathcal V(t_2,t_1)$ is the Markovian map generated by Eq.~\eqref{eq:Lindblad_equation} in the main text, ${\mathcal V}_{\rm st}(\xi,t_2,t_1)$ is a unitary trajectory map depending on either a realization of the discrete stochastic field $\xi_{kn}$ (EC1 and EC3) or a random choice of unitaries (EC2).
We use the Dyson-like expansion in Eq.~\eqref{eq:V_sp_decomposition2} and the convexity of the absolute value to upper bound the error as
\be
\begin{split}
\epsilon\leq & \frac {\Delta\tau}2 \mathbb E\max_{\vec r'}\int_0^t dt'\sum_{\vec r}\Bigl|\<\psi_{\vec r}|\mathcal V_{\rm st}(t,t') \mathcal D(t')\mathcal V_{\rm st}(t',0)\rho_{\vec r'}|\psi_{\vec r}\>\Bigl|\\
&+O(\Delta\tau^2).
\end{split}
\ee
Using the fact that $\mathcal V_{\rm st}$ is a unitary map, we can rewrite
\be
\mathcal V_{\rm st}(t',0) = \mathcal V^{-1}_{\rm st}(t',t)\mathcal V_{\rm st}(t,0),
\ee
where the inverse of a unitary map is well defined through the inverse unitary transformations. This expression leads directly to Eq.~\eqref{eq:erro_int_step1}, taking into account that
\be
\mathcal V^{-1}_{\rm st}(t',t)\mathcal D(t)\rho \mathcal V^{-1}_{\rm st}(t',t) = \sum_\alpha D^{(1)}_{\alpha}(t,t')\rho D^{(2)}_{\alpha}(t,t'),
\ee
where $D_\alpha^{(i)}(t,t') = \mathcal V_{\rm st}(t,t') D_\alpha^{(i)}(t')$.

\subsection{Proof of the lemma}

Let us rewrite the left-hand side of Eq.~\eqref{eq:lemma_statement} using the spectral decomposition $\rho = \sum_\mu p_\mu |\phi_\mu\>\<\phi_\mu|$ and triangle inequality as
\be
\begin{split}
\sum_{\rm \vec r}& |\<\psi_{\vec r}|O_1\rho O_2|\psi_{\vec r}\>| = \sum_{\vec r} \Bigl|\sum_{\mu,\vec r_1 \vec r_2}p_\mu\phi^\mu_{\vec r_1}\phi^{\mu*}_{\vec r_2}\<\psi_{\vec r}|O|\psi_{\vec r_1}\>\<\psi_{\vec r_2}|O|\psi_{\vec r}\>\Bigl| \\
&\leq \sum_\mu p_\mu \sum_{\vec r}\sum_{\vec r_1 \vec r_2}|\phi^\mu_{\vec r_1}||\phi^\mu_{\vec r_2}||\<\psi_{\vec r}|O|\psi_{\vec r_1}\>||\<\psi_{\vec r_2}|O|\psi_{\vec r}\>|\\
&\leq \|O_1\|_{\rm max}\|O_2\|_{\rm max} \sum_{\mu, \vec r}\sum_{\vec r_1\in D(k_1,\vec r)}\sum_{\vec r_2\in D(k_2,\vec r)} p_\mu|\phi^\mu_{\vec r_1}||\phi^\mu_{\vec r_2}|,
\end{split}
\ee
where we denote $\|O\|_{\rm max} = \max_{ij} |O_{ij}|$ to be the max-norm of the matrix $O$, and $D_k(\vec r)$ is a sphere with radius $k$ with respect to Hamming distance.
Using the inequality
\be
|\phi^\mu_{\vec r_1}||\phi^\mu_{\vec r_2}| \leq \frac 12\Bigl(|\phi^\mu_{\vec r_1}|^2+|\phi^\mu_{\vec r_2}|^2\Bigl) 
\ee
and the property that the sphere $D(k,\vec r)$ contains $\binom{L}{k} \leq L^k/k!$ states, we obtain
\be
\sum_{\rm \vec r}|\<\psi_{\vec r}|O_1\rho O_2|\psi_{\vec r}\>| \leq \frac 1{k_1!k_2!}\|O_1\|_{\rm max}\|O_2\|_{\rm max}L^{k_1+k_2},
\ee
where we use the fact that the density matrix is properly normalized, $\Tr\rho=1$.\\

\section{Dissipative gates errors \label{app:gates}}

In this Appendix, we derive the error of dissipative gates analyzed in Sections \ref{sec:hard} and \ref{sec:atoms}. 

In the case of imperfect Zeno blockade, the major source of error is associated with leakage to the out-of-logic states. In the scheme proposed in Section \ref{sec:hard}, there are two relevant out-of-logic states into which leakage occurs from the state $|00\>_L = |0101\>$, namely $|0011\>$ and $|0000\>$. The first of these states ($|0011\>$) is accessed via a unitary channel,  while the second of these states ($|0000\>$) is accessed via a dissipative channel. The simplest way to describe leakage is to consider unitary evolution of basis states $\{|0101\>,|0011\>\}$ and including the second-channel leakage using a non-Hermitian term. The resulting non-Hermitian Hamiltonian is
\be
H_{S}=\left(
  \begin{matrix}
    0 & J  \\
    J & -\frac{i}2\Gamma
  \end{matrix}\right).
\ee

The leakage error can be computed as
\be
\begin{split}
|\<00|_LS|00\>_L|^2 \equiv 1-2\epsilon
= 1-\frac{4\pi}{\gamma}+O\Bigl(\frac{1}{\gamma^2}\Bigl),
\end{split}
\ee\textbf{}
where $S = \exp(-i\pi H_S/J)$ and $\gamma = \Gamma/J$.

For the scheme involving cold atoms in Section \ref{sec:atoms}, the relevant out-of-logic states are
  \be\label{out_of_logic_states2}
 \begin{split}
 &|P1\> = |0,\mu_2\mu_1,0,0\>,\\
 &|P2\> = |0,0,\mu_1\mu_2,0\>,\\
 &|EX\> = |0,\mu_2,\mu_1,0\>,\\
 &|VC\> = |0,0,0,0\>.
 \end{split}
 \ee
The restriction of the effective Hamiltonian to the subspace spanned by the basis $\{|01\>,|P1\>,|P2\>,|EX\>\}$ is
\be
H'_{S}=\left(
  \begin{matrix}
    0 & J & J & 0 \\
    J & E-\frac{i}2\Gamma & 0 & J \\
    J & 0 & E-\frac{i}2\Gamma & J \\
    0 & J & J & 0 \\
  \end{matrix}\right),
\ee
where $E$ is the interaction energy of a fermion pair on the same site. The non-Hermitian nature of the Hamiltonian reflects additional leakage to the fully empty state $|VC\>$ due to pair loss. In the strong dissipation limit $J\ll \Gamma$, the leakage error $\epsilon$ for the gate can be defined as
\be\label{eq:zeno_error}
\begin{split}
1-2\epsilon&\equiv |\<01|_LS'|01\>_L|^2\\
&= 1-\frac{8\pi^2}{\Gamma t}\frac 1{1+4\zeta^2}+O\Bigl(\frac{1}{\Gamma^2t^2}\Bigl),
\end{split}
\ee
where $S' = \exp(-iH'_{S}t)$, $t=\pi/J$ is the characteristic time of hopping between lattice sites, and $\zeta = E/\Gamma$ is the ratio between the interaction energy and the pair loss rate.

\bibliography{library}

\end{document}